\documentclass[referee]{aa}  

\usepackage{graphicx}
\usepackage{txfonts}
\usepackage{lipsum}
\usepackage{subcaption}         
\usepackage{lscape}             
\usepackage{placeins}           
                                
\begin{document}

   \title{Context images for Venus Express radio occultation measurements: A search for a correlation between temperature structure and UV contrasts in the clouds of Venus}

   \author{
      M. Roos-Serote\inst{1,2}
      C.F. Wilson\inst{2,3}
      R.J. MacDonald\inst{2,4}
      S. Tellmann\inst{5}
      Y.J. Lee\inst{6}
      I.V. Khatuntsev\inst{7}
    }

   \institute{Lightcurve Films, Portugal\\
             \email{science@lightcurvefilms.com}
	\and AOPP, University of Oxford, UK 
	\and European Space Agency, ESTEC, Noordwijk, the Netherlands
	\and Department of Astronomy, University of Michigan, Ann Arbor, United States
	\and Rheinisches Institut für Umweltforschung, Universität zu Köln, Germany
	\and Planetary Atmospheres Group, Institute for Basic Science (IBS), Daejeon, South Korea
	\and Space Research Institute of the Russian Academy of Sciences (IKI), Moscow, Russia
       }

   \date{}

    \titlerunning{Correlation between temperature and UV-contrast in Venus' clouds}
    \authorrunning{Roos-Serote et al.}
    
 
  \abstract
   {Venus exhibits strong and changing contrasts at ultraviolet wavelengths apparently related to the clouds and the dynamics in the cloud layer, but to date their origin continues to be unknown.}
   {We investigate the nature of the UV contrasts exhibited by Venus’ clouds by examining possible correlations between the thermal structure inferred from radio occultation data and UV brightness from imagery data, both observed with Venus Express.} 
   {We analyse Venus Express images obtained from 11 hours before to a few hours after the time of radio occultation measurements of the same area.
   We account for the advection of clouds by zonal and meridional winds and apply a phase angle correction to compensate for the changing viewing geometry.}
   { 
    We find a possible anti-correlation between UV-brightness and atmospheric temperature in the 65-70 km altitude range for low latitudes. 
    Heating in this altitude and latitude region due to an increase in the UV-absorber has been predicted by radiative forcing studies.
    The predictions roughly match our observed temperature amplitude between UV-dark and UV-bright regions.
    We find no evidence for any correlation between UV-brightness and static stability in the atmosphere in the 50-80 km altitude region.     
    }
   {This could be the first observational evidence for a direct link between UV-brightness and atmospheric temperature in the 65-70km altitude region in the clouds of Venus.}


   \maketitle

\section{Introduction}
\label{introduction}

Venus is completely enveloped by highly reflective clouds.
Based on analyses of their phase functions and spectra, the upper clouds are known to be composed principally of sulphuric acid.
At visible and infrared wavelengths the Venus clouds look very homogeneous.
However there are strong contrasts at UV wavelengths, which were first observed early in the early twentieth century \citep{Wright1927PASP, Ross1928ApJ}. 
\citet{Ross1928ApJ} suggested that these UV contrasts might be correlated to temperature differences between the north and south half of Venus, 
which he based on temperature estimates derived from broad-band IR emission measurements reported by \citet{Coblentz1925}.
Significant UV absorption at wavelengths below 300 nm is due to sulphur dioxide and ozone at latitudes poleward of $\pm$ 50$^{\circ}$ \citep{Marcq2019Icarus},
but an additional UV-to-blue absorber, or absorbers, is required at 300 – 450 nm to match the observed spectrum \citep{Pollack1980JGR, Marcq2020Icarus}.
The chemical identity of this UV-to-blue-absorber is still unknown.
Proposed compositions include a mixture of elemental sulphur allotropes \citep{Toon1982Icarus},
ferric chloride \citep{Esposito1997VenusII} and disulphur dioxide \citep{Frandsen2016GRL}.
Concerning the vertical distribution, measurements from descent probes concur that the UV-to-blue absorber lies above 57 km \citep{Tomasko1980JGR, Ekonomov1984Nature},
but that the peak absorption must be near but below the cloud top level in the 66-73 km altitude range \citep{Crisp1986Icarus, Lee2015PSS} to explain the observed phase angle dependence of the contrast \citep{Pollack1980JGR}
and the mean phase curve of the disk brightness \citep{Lee2021GRL}.

Just as the identity of the UV-to-blue-absorber is unknown, the reasons for the UV contrasts are also unknown. \citet{Titov2008Nature} report that UV contrast changes are associated with upwelling from below, and not with changes in the cloud top altitude, inferred from data of the Venus Express / VIRTIS instrument.
However, it is not clear whether the upwelling advects UV-dark material such as poly sulphur or FeCl$_3$, or UV-bright material such as sulphuric acid haze particles, which are more spatially variable.
If the contrasts are associated with upwelling in the cloud layer, is this upwelling bringing UV-dark material or UV-bright haze-forming material to the cloud-tops?
\citet{Cottini2015PSS} present a search for correlations between the cloud top altitude measured at 2.5 micron wavelength from Venus Express VIRTIS-H spectra, and the UV-brightness measured simultaneously by VIRTIS-M between 375 nm and 385 nm.
They report that there might be some anti-correlation with darker UV areas corresponding to higher cloud tops (denser clouds), but it is not systematic (their Fig. 8).
\citet{Patsaeva2015PSS} used altimetry from VIRTIS-M for VMC UV images and showed that cloud motion vectors obtained for darker regions lying 1-1.5 km above adjacent bright regions give a different horizontal flow direction (their Fig. 11).
\citet{Lee2020Nature} report on a clear anti-correlation between the overall brightness of Venus at 283 nm (SO2 absorption) and 365 nm (unknown absorber) and at 2.02 micron in several years of Akatsuki data. The 2.02 micron wavelength area is very sensitive to cloud top height, because it is strongly affected by carbon-dioxide absorption.

These are mechanisms by which temperature contrasts and dynamics could lead to changes in the UV-brightness.
However, the UV-to-blue-absorber also has radiative effects; in particular, UV-dark regions should be expected to absorb more sunlight, at least in the UV part of the spectrum, and thus might be expected to experience higher temperatures than UV-bright regions.
The present analysis of co-located UV imagery and radio occultation measurements could provide constraints on the vertical distribution of UV absorption and validation of previous calculations of UV-contrast related heating rates 
\citep{Crisp1986Icarus, Lee2019AJ}.

The Venus Express mission \citep{Svedhem2007PSS} produced very useful and unique datasets.
The analysis of this data can help answer the questions above.
In particular, with the VEnus RAdio science (VeRa) experiment radio occultation is used to obtain vertical density profiles of the atmosphere with high vertical resolution from which, by integration and assumption of hydrostatic balance, the pressure and temperature profiles can be calculated \citep{Tellmann2009JGR}.
The cloud contrasts are most easily observed in images obtained with the UV channel of the Venus Monitoring Camera (VMC), which has a bandpass centred at a wavelength of 365 nm, at the peak of the unknown UV absorption \citep{Markiewicz2007PSS}.
The Visible and InfraRed Thermal Imaging Spectrometer (VIRTIS) instrument obtains hyperspectral images which include coverage of the UV feature, but are not favoured in this analysis for two reasons: first there are many fewer UV VIRTIS images than there are UV VMC images, and they cover a smaller area due to the smaller field of view; second, the VIRTIS UV-VIS channel suffers from stray light problems at wavelengths below 500 nm which are not yet fully resolved.
On the other hand, VIRTIS-IR observations offer two relevant observational constraints on cloud processes: (1) dayside observations of CO$_2$ line depths measure the height of the cloud-top at these wavelengths \citep{Ignatiev2009JGR};
and (2) nightside observations in 1.7 and 2.3 $\mu$m windows offer the possibility to constrain lower cloud optical thicknesses \citep{Barstow2012Icarus}.
 
In the present study, we identify orbits during which both VeRa radio occultation sounding and VMC imaging were done.
The analysis of this data allows us to look for correlations between the VeRa temperature profile and static stability with UV-brightness as measured by VMC.
The observations and analysis procedure are described in Section \ref{dataandreduction}, followed by an analysis of the results and a discussion in Section \ref{analysis}.
We did assess the possibility of studying correlations using VIRTIS nightside imagery, but too few co-located observations were found for a meaningful analysis to be done.

\section{Data and reduction}
\label{dataandreduction}

\subsection{Selection of observations}
\label{selectionofobservations}

Venus Express was equipped with two High Gain antennas, pointing in directions nearly aligned with the spacecraft’s +X and –X axes respectively.
The VMC instrument boresights on the other hand were directed in the +Z direction.
Hence there could never be any observations of the same location by VMC and VeRa simultaneously.
To obtain imagery of a VeRa sounding location it would have been necessary to slew the spacecraft before or after the VeRa sounding.
For observations of the Northern hemisphere the spacecraft’s altitude above the clouds was too low, and hence the spacecraft velocity too high, to make this feasible.
It was only possible for observations of the Southern Hemisphere when the spacecraft was much further from the planet on its elliptical orbit and at much lower orbital velocities.

  \begin{figure}[h!]
   \centering
   \includegraphics[width=\hsize]{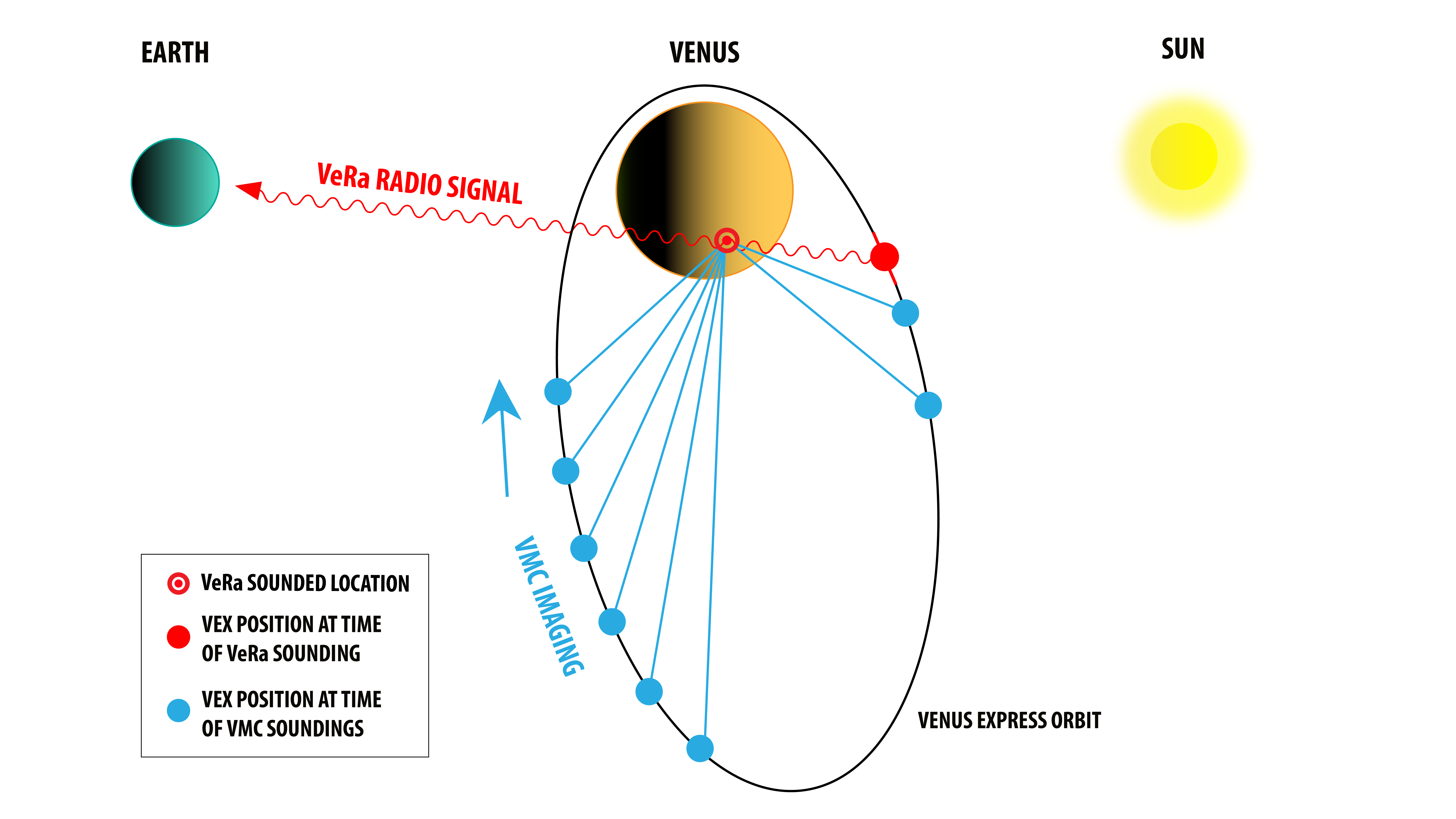}
      \caption{Schematic of Venus Express orbit around Venus with VMC-imaging happening on ingress and egress (blue dots) and VeRa radio occultation shortly after periapsis (red dot).}
         \label{figure1}
   \end{figure}
 
A dedicated South Polar Dynamics Campaign (SPDC) was carried out between 25 November and 31 December 2013, specifically designed to allow imagery of VeRa sounding locations (orbits numbers 2775 through 2811, last section in Table \ref{table1}).
On each orbit one VeRa atmospheric sounding was acquired shortly after the pericentre passage, sounding high Southern latitudes (-48° to -83°).
Before and after the VeRa sounding the VMC camera boresight was pointed at the planet and images were taken.
On ingress VMC acquired images at half-hourly intervals over a period of several hours.
The geometry of the observations is shown schematically in Fig.~\ref{figure1}.
After pericentre a brief period of about one hour was available for imaging.
For 30 orbits from the SPDC good data from both VeRa and VMC are available within a time separation of less than 12 hours of each other.

\begin{figure}[h!]
   \centering
   \includegraphics[width=\hsize]{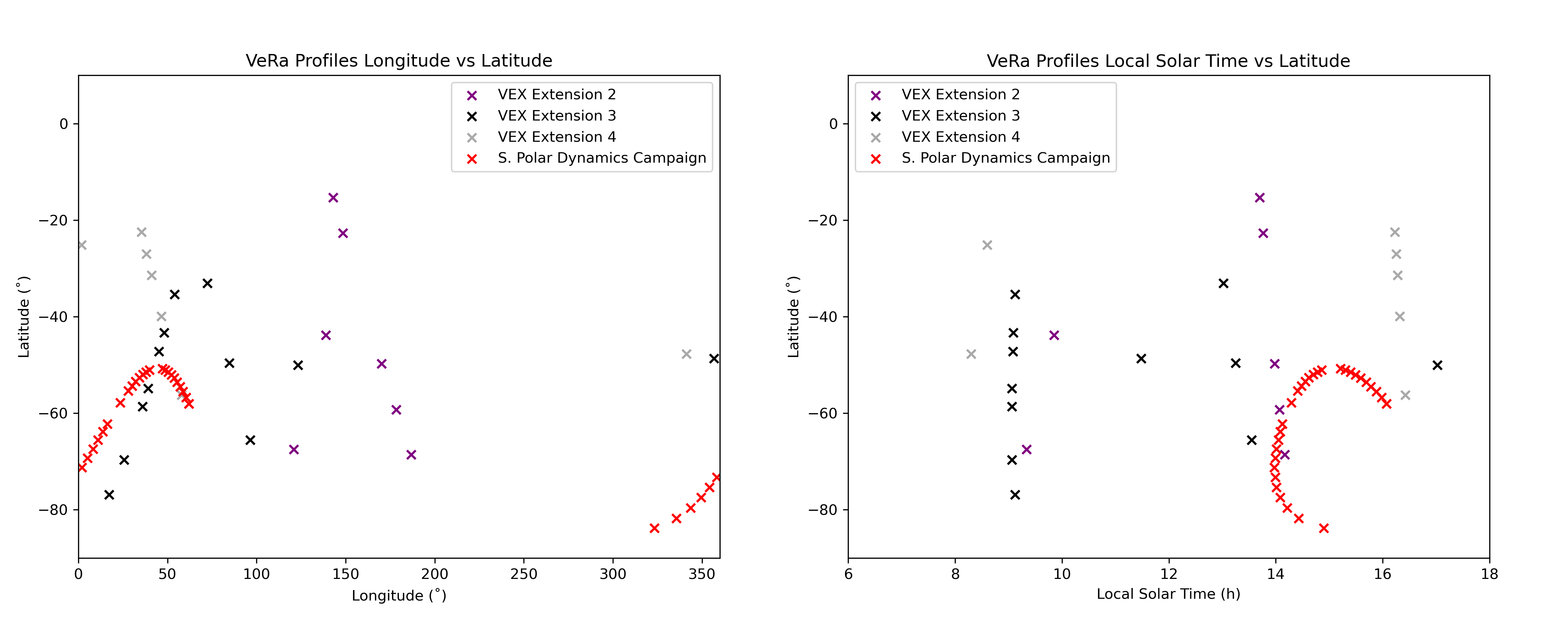}
      \caption{Location on Venus of all the VeRa radio occultation soundings included in this study in terms of latitude - longitude (left) and latitude - Local Solar Time (right).}
         \label{figure2}
   \end{figure}

During the rest of the Venus Express mission VMC images of Venus that contain VeRa sounding locations within a few hours of the sounding had not been planned, but do occur.
We identified an additional 26 orbits during which co-located VeRa and VMC data are available and usable for analysis.
These cover the VEX mission extensions 2 through 4, starting at orbit number 1191.
Including these observations allows for our analysis to extend to latitudes down to the low Southern latitudes.
All the selected observations are listed in Table \ref{table1} and the distribution of the VeRa profiles on Venus is shown in Fig.~\ref{figure2} as a function of latitude, longitude and Local Solar Time.

\subsection{VeRa dataset}
\label{veradataset}

The VeRa data has been described in detail in \citet{Patzold2007Nature} and \citet{Tellmann2009JGR}.
The processed VeRa datasets consist of density, pressure and temperature profiles of the neutral atmosphere from approximately 50 to 100 km altitude.
The vertical resolution of VeRa temperature profiles depends on observation geometry, but is typically on the order of several hundred metres.
For our analysis we apply a binning scheme for each profile: we  calculate the average value of the VeRa temperatures and pressures inside 1-km bins between 46 and 102 km altitude.
This is equivalent to applying a low pass filter,  and also helps to minimise any effects of gravity waves on our data and consequently our analysis. We discuss gravity waves in more detail below.

\begin{figure}[h!]
   \centering
   \includegraphics[width=\hsize]{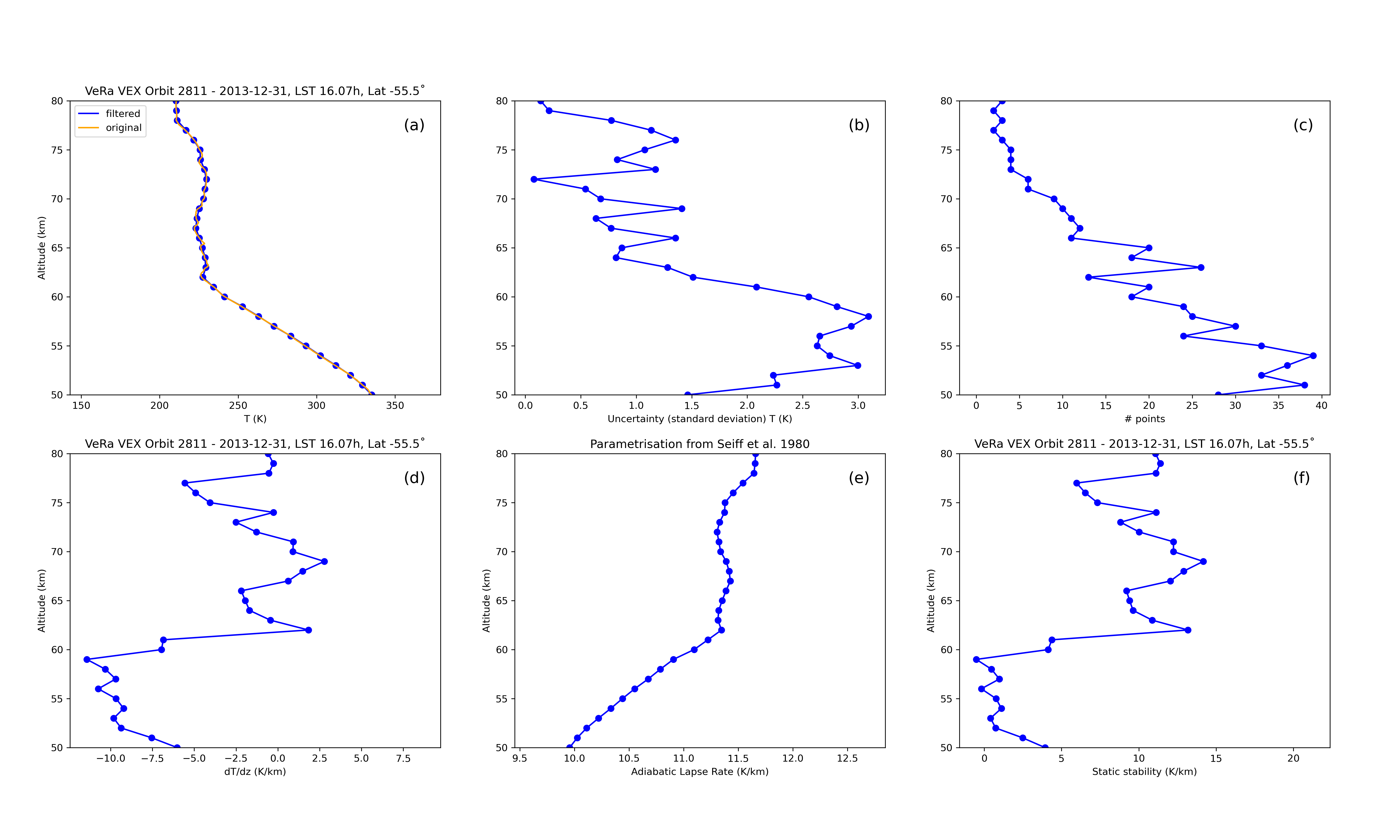}
      \caption{Example of a temperature profile derived from VeRa radio occultation data for orbit ID 2811 (31 December 2013). (a) the 1-km binned profile (dots) and the original profile (line). Binning removes the high frequency components (see text); (b) the standard deviation of the temperature values that were averaged in each altitude bin, which is an indication of the uncertainty of the binned temperature values; (c) the number of values in each altitude bin; (d) temperature gradient derived from the temperature profile; (e) adiabatic lapse rate from \citet{Seiff1980JGR}; (f) static stability is the temperature gradient minus the adiabatic lapse rate.}
         \label{figure3a-f}
   \end{figure}

In Fig.~\ref{figure3a-f}a-c we show an example of such a binned profile, compared  to its full profile.
We also compute the temperature gradient (Fig.~\ref{figure3a-f}d) and static stability profiles S(z) for each temperature profile  (Fig.~\ref{figure3a-f}f).
Static stability is defined as the observed temperature gradient rate minus the adiabatic lapse rate: we derive the adiabatic lapse rates from Figure 18 in \citet{Seiff1980JGR} (Fig.~\ref{figure3a-f}e) and calculate S(z) at 1 km intervals, as shown in Fig.~\ref{figure3a-f}f.
A sharp change in regime can be seen from a low stability indicating convective overturning in the lower cloud deck between 50 and 60 km altitude, to a high stability upper cloud layer extending upwards from 60 km altitude.
In the present study, we look for correlations between the UV-brightness and both temperature and static stability at all relevant altitudes.

Particular care should be taken with analysing radio occultation temperature data when strong inversions are present, which can occur at high latitudes near the tropopause around 60 km.
In this case the radio signal may take multiple paths through the atmosphere, causing an error in the retrieved temperature profile of  $\approx$5K in magnitude \citep{Herrmann2014EPSC}.
Corrections for multi-path effects are not routinely available at the time of writing, and so were not used in the present analysis.
We note that it may introduce errors in the temperature profile in the 58–62 km altitude range, at high latitudes poleward of $\pm$60°.
The altitudes targeted in this analysis are 65km and higher, and 17 out of 56 profiles assessed in this work were obtained at latitudes higher than -60° (Table \ref{table1}).

There are also known effects on the temperature structure at and above the cloud level due to thermal tides and gravity waves.
These effects have been studied by many authors with data from different Venus missions. 
\citet{Zasova2007PSS} and \citet{Imamura2018JGR} present analysis of Venus Express data, \citet{Kouyama2017GRL}, \citet{Imamura2018JGR} and \citet{Akiba2021JGR} present analysis of Akatsuki data.
Thermal tides are a function of latitude and Local Solar Time.
\citet{Akiba2021JGR} present amplitudes of these tides as a function of latitude and altitude.
The altitude range however is limited between 65 and 72 km.
Our analysis extends between 50 and 80 km altitude.
The reported thermal tide amplitudes are mostly in the 1-3 K range, which is on the order of the uncertainties in the temperature values in our analysis (Fig.~\ref{figure3a-f}b).
In order to evaluate the effect of applying a thermal tide correction we took the available data for Figure 5 in \citet{Akiba2021JGR} from the corresponding Zenodo-repository.
This data is for the 69 km altitude level.
We extracted thermal tide amplitudes for the latitudes and Local Solar Times of the VeRa soundings that we used.
We performed the statistical analysis, described in detail in Section \ref{analysis}, without and with the correction for the thermal tide.
The resulting values for the Pearson and Spearman correlation coefficients for the three latitude bins that we defined (low, mid and high latitude, see details in Section \ref{analysis}) are the same within the uncertainties.
Hence, we decide that attempting to correct for the thermal tide is not necessary for the data we use and the way we analyse it.

The exact origin of gravity waves in the atmosphere of Venus is still under debate.
\citet{Tellmann2012Icarus} report on small scale temperature fluctuations of up to $\pm$4K as derived from analysis of VeRa data from all over Venus.
They observe the onset of wavelike temperature fluctuations at around 60 km altitude, the tropopause in the middle of the cloud layers.
Here the atmosphere changes from an adiabatic to convectively stable state, with excursions back to adiabatic in thin confined layers throughout the cloud layer.
The amplitude of the temperature fluctuations diminishes with altitude all the way up to 90 km.
Towards the latitudes of $\pm$60$^{\circ}$, the polar collar regions, the overall amplitudes of the temperature fluctuations are larger.
At this latitude the polar collar is observed, with much lower tropopause temperatures, a phenomenon not yet fully understood.
\citet{Imamura2018JGR} re-analyse Venus Express data and Akatsuki data using the so-called Full Spectral Inversion method for the first time on this type of radio occultation data.
This technique allows to account for multi-path rays.
The resulting vertical resolution is about seven times higher (150m) when compared to the classically used Geometric Optics method which has been applied in VeRa analysis.
They also find the same trend as observed by \citet{Tellmann2012Icarus}  at the 1-km vertical resolution level.
Thanks to the higher vertical resolution, more defined thin adiabatic layers throughout the upper part of the cloud layers can be seen.
 
\citet{Kouyama2017GRL} report on the discovery of semi-stationary gravity waves linked to four topographic regions with mountains.
These waves are seen in the form of maxima in the thermal infrared temperatures with an amplitude of up to $\pm$1.5K.
They are semi-stationary in longitude and appear most strongly when the topographic region is in the local afternoon, and disappear during the night.
 
Gravity waves can occur anywhere in the atmosphere and are not reported to have a local time dependence.
As mentioned earlier in this paragraph, we have applied averaging in 1-km wide bins, which acts as a low pass filter and minimises the effects of gravity waves on our analysis.

\subsection{VMC dataset}
\label{vmcdataset}

To perform the VMC – VeRA correlation, two steps were undertaken.
First, the movement of the atmosphere between the time of a VMC image and a VeRa radio occultation sounding due to winds was taken into account. 
Second, the instrument calibration and illumination conditions were taken into account to calculate a phase-corrected relative brightness.

\subsubsection{Correction for wind advection}
\label{correctionforwindadvection}

The time difference between VMC and VeRa observations $t_{VeRa}$ - $t_{VMC}$  ranged from  –11 to +1 hours.
In this time interval the cloud field is advected by atmospheric winds, which needs to be taken into account when comparing the observations.
We evaluate two possible approaches for this correction, using mean wind fields, and orbit-specific wind fields.
 
The most comprehensive analysis of mean zonal and meridional wind speeds for cloud-top level using VMC-UV data is that reported by \citet{Khatuntsev2013Icarus}.
In Figure 10 of their paper the average zonal and meridional winds are shown as a function of latitude between 0º and -80º latitude.
The strong zonal wind slowly increases from about -90 m/s at the equator to -100 m/s at -46º latitude. It drops towards 0 m/s at the pole.
The meridional wind varies between 0 m/s near the equator and pole to about -10 m/s around -45º latitude.
The figure also shows the dispersion in measured wind speeds:  $\pm$20 m/s for the zonal wind and $\pm$12.5 m/s for the meridional wind.
To access the wind speeds at any latitude we use a linear parametrisation of the curves in latitude sections, making sure to stay within the error bars.
For the zonal wind we parametrise for latitude sections between [-90$^{\circ}$, -50º], [-50º, -40$^{\circ}$], [-40º, -15$^{\circ}$] and latitudes > -15º.
For the meridional wind the latitude sections are [-90$^{\circ}$, -75º], [-75$^{\circ}$, -50$^{\circ}$], [-50º, -20$^{\circ}$] and latitudes > -20º.
Longitudinal or Local Solar Time variation of the mean winds were not taken into account.
 
The next step is to determine where the parcel of atmosphere sounded during a VeRa radio occultation at location ($\phi_{VeRa}$, $\theta_{VeRa}$) has advected to at the time a VMC image is acquired.
We determine the coordinates of this point, ($\phi_{u-average}$, $\theta_{Vu-average}$), from the time difference between the VMC image and the VeRa sounding combined with the average wind speeds at the latitude $\theta_{VeRa}$, both in the zonal and meridional direction.
We evaluate the uncertainty in ($\phi_{u-average}$, $\theta_{Vu-average}$), by doing the same calculation using the spread in the zonal ($\pm$ 20 m/s) and meridional ($\pm$ 12 m/s) winds.
We thus obtain four additional coordinates that define a latitude / longitude box centred at ($\phi_{u-average}$, $\theta_{Vu-average}$).
The size of this box is about 30º in longitude and 5º in latitude at the largest time difference of -11h, and about one tenth of this at the smallest time difference of plus one hour.
 
The final step is to determine the average of the UV-brightness within this box and take this as the VMC radiance for the sounding location for that image.
The UV-brightness of each pixel is calculated taking into account the incidence and phase angles of the observation, as described in Section \ref{calculationofrelativeuvbrightness}.
The uncertainty is taken to be the standard deviation of the average of all the pixels within the latitude-longitude box described above.
 
The winds at any given time may deviate from their mean values.
We evaluated whether when using wind speeds values obtained during a specific orbit a more accurate advection correction would be obtained, when compared to using mission mean wind speeds.
This was particularly possible for the data from the South Polar Dynamics Campaign, because VMC images were taken at regular intervals (about 30 minute) intervals.
\citet{Khatuntsev2024Mendeley} provided  with wind vectors for 18 orbits from the South Polar Dynamics Campaign.
When comparing the centres of the latitude-longitude boxes derived in this way, with those derived when using mission average wind speed values, they overlap within the uncertainties.
In fact, the uncertainties in the positions when using orbit specific wind speed values were found to be higher.
We therefore decided it to be more consistent to use the mission average values throughout our analysis.
 
   \begin{figure}[h!]
   \centering
   \includegraphics[width=\hsize]{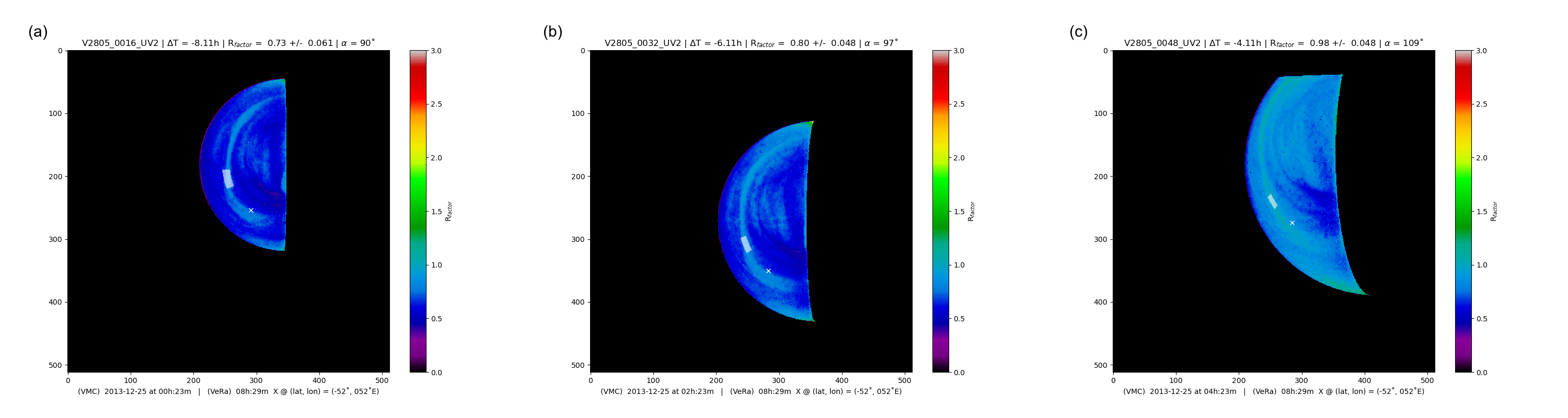}
      \caption{Three images taken two hours apart on the ingress of orbit 2805 on 25 December 2013. The cross indicates the spot of the VeRa radio occultation that happens after 8h, 6h and 4h respectively. The rectangle is the wind-advected latitude / longitude box corresponding to VeRa location advected by the zonal and meridional winds. The boundaries of the box are a measure of the the uncertainties in the zonal and meridional winds (see main text for details). It can be seen how the rectangle becomes smaller when closer in time (and hence in space) to the VeRa sounding location.}
         \label{figure4a-c}
   \end{figure}

In Fig.~\ref{figure4a-c}a-c we illustrate the result of the procedure showing a set of images from orbit 2805 (25 December 2013) obtained 8h, 6h and 4h before the VeRa sounding on that orbit.
In each figure, the latitude-longitude position of the VeRa sounded location is marked by a cross, the predicted location of the sounded air parcel ($\phi_{u-average}$, $\theta_{Vu-average}$) is marked as a grey square.
It can be seen that the square stays roughly in the same position with respect to cloud features in the time span between the VMC and VeRa observations, illustrating that the mean wind field captures the actual cloud motion accurately enough.

\subsubsection{Calculation of relative UV-brightness}
\label{calculationofrelativeuvbrightness}

The VMC data used for this analysis are calibrated level 2 images obtained from the ESA Planetary Science Archive.
The radiances measured in the images are a function of the incidence, emission and phase angles of the observation.
We applied a simple Lambert disk function as photometric correction: the radiance divided by the cosine of the incidence angle.
\citet{Lee2015Icarus} demonstrate that this is sufficiently accurate (their paragraph 3.3) for viewing geometries where the phase angle does not exceed 130$^{\circ}$ and the incidence and emission angles are smaller than 84$^{\circ}$ and 81$^{\circ}$ respectively.
We apply these same constraints for the images in our analysis.
 
\begin{figure}[h!]
   \centering
   \includegraphics[width=\hsize]{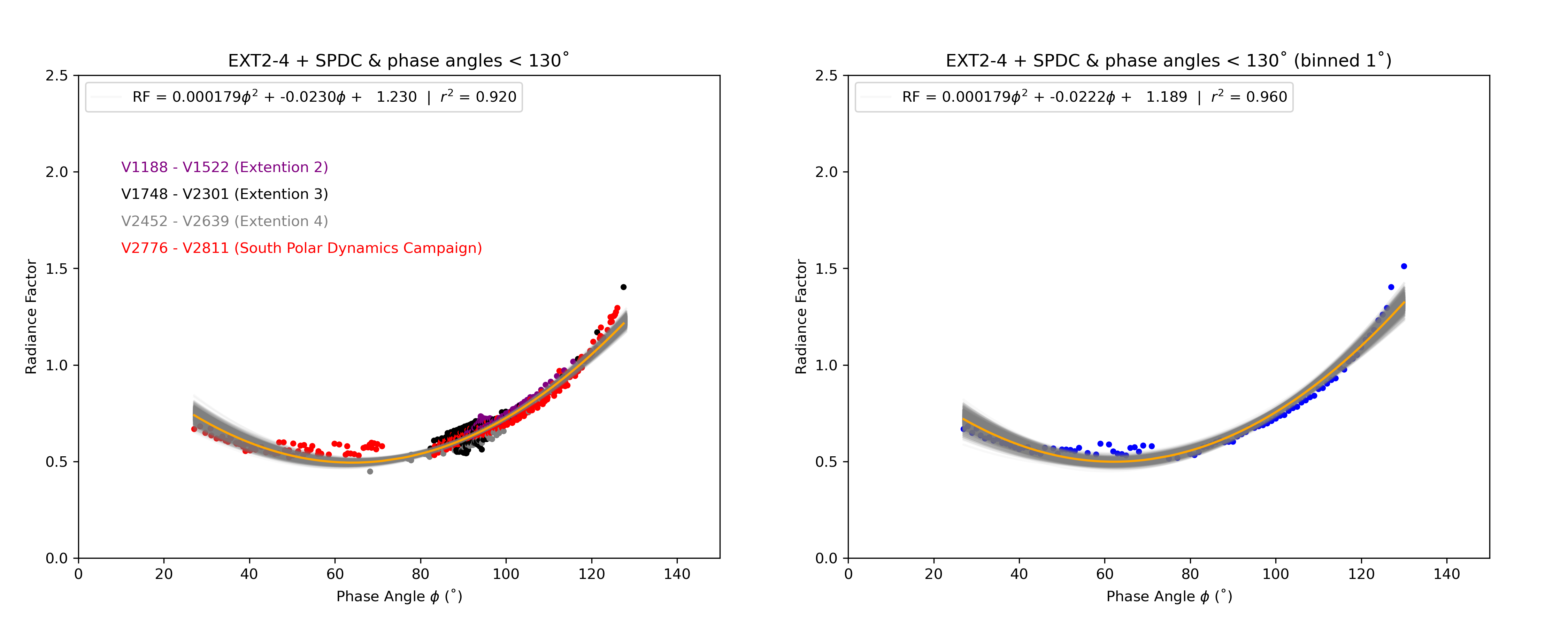}
      \caption{The phase curve for all valid images of this study: (a) all images individually (b) same as (a) but binned in bins of 1$^{\circ}$ width and averaged in the bins. The orange curve is the best quadratic fit to the points and grey bands are a measure for the uncertainties on the curve.}
         \label{figure5a-b}
   \end{figure}

In order to be able to compare radiances retrieved from images taken at different phase angles, we must take into account the scattering phase function.
For this we build a phase curve making use of all the 972 images from the 56 orbits in our analysis.
We calculate the average Radiance Factor \citep[see][eq.~2]{Lee2015Icarus} for each image taking all the pixels with the geometry conditions as described in the previous paragraph.
In Fig.~\ref{figure5a-b}a we present the phase curve with this data, and the quadratic least square fit.
We found that binning the results in phase angle bins of 1$^{\circ}$ width and taking the average in each bin results in an improved fit with a higher confidence level, as can bee seen in Fig.~\ref{figure5a-b}b. 

The grey area around the model curves in Fig.~\ref{figure5a-b} is the uncertainty range.
We derived this uncertainty range by randomly varying each of the bin-averaged Radiance Factors.
To do this we add a randomly drawn value from a Gaussian distribution to each bin-averaged Radiance Factor value.
The average of this distribution is the bin-averaged Radiance Factor value, the standard deviation is the maximum of all the uncertainties on the individual Radiance Factors values in the bin and the uncertainty as assessed through equation 3.14 from \citet{Bevington2003}.
We do this exercise 1000 times and for each of these permutations we fit a new model.
We verify that the average Radiance Factor value of the models of the 1000 experiments for each phase angle bin is within 0.5\% of the actual bin-averaged Radiance Factor.
There are two ways to derive the uncertainty of the model Radiance Factor in each bin: either take the standard deviation of the 1000 experiments or take half the value of the maximum minus the minimum Radiance Factors of the 1000 experiments.
We compare both and decide to take the maximum of the two, which turns out to be the second option and is on the order of 10\% for each bin Radiance Factor.
 
\begin{figure}[h!]
   \centering
   \includegraphics[width=\hsize]{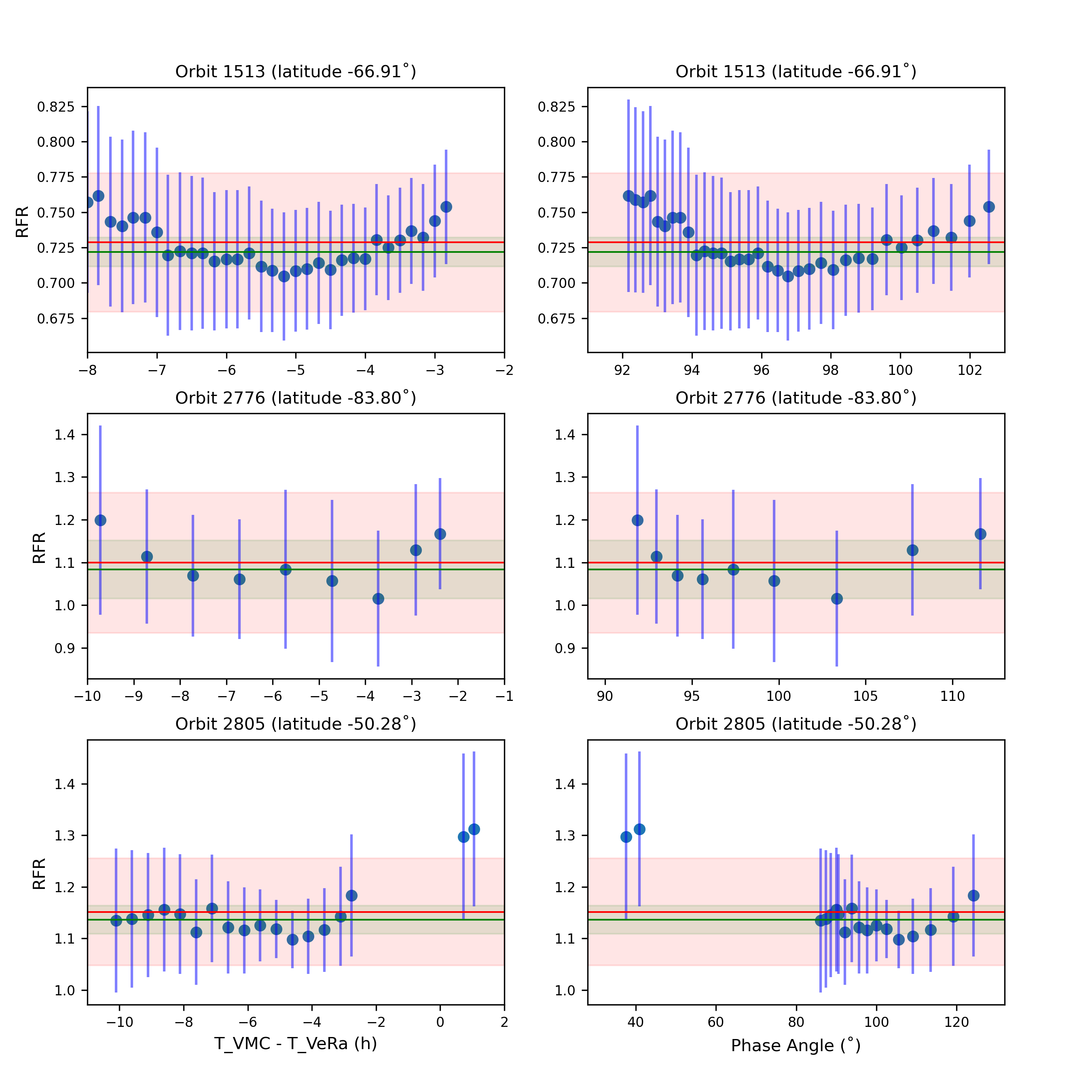}
      \caption{Examples of the Radiance Factor Ratios for sequences of images on a same orbit, as derived from the radiance factors extracted from the latitude / longitude boxes of the wind-advected areas (Fig.~\ref{figure4a-c}) and applying the phase curve model (Fig.~\ref{figure5a-b}). We expect these values to be very similar throughout one orbit. The average and standard deviation is shown in red, the median and associated uncertainties in green. See also Table~\ref{table1}.}
         \label{figure6}
   \end{figure}

As presented in Section \ref{correctionforwindadvection} we track the area of a VeRa sounding (latitude / longitude boxes) in the different images taken during the orbit.
Given the scale of the latitude / longitude boxes and the uncertainties involved (Fig.~\ref{figure4a-c}a-c), the average Radiance Factors of latitude / longitude boxed measured from each of the images of one orbit should be similar.
The observing geometry changes between images on an orbit.
Hence, to compare between images on one orbit we use the model phase curve to normalise the average Radiance Factors for each image.
These normalised Radiance Factors we call Radiance Factor Ratios (RFR).
In Fig.~\ref{figure6} we present examples of RFR for the tracked VeRa areas on three different orbits, as a function of the time difference between the VMC image and the VeRa sounding (left column) and the phase angle (right column).
The last row is for orbit 2805, the same as the example images shown in Fig.~\ref{figure4a-c}.
The red line is the average RFR-value over all the images of the orbit, the associated uncertainty (standard deviation) is represented by the red-shaded area.
The green line and green-shared area represent the median values and associated uncertainties (see Table~\ref{table1}).
The RFR-values per image on one orbit are consistent taking into account their uncertainties (blue error bars).

We recognise that our assumption of the phase function being applicable over all regions of the planet, including both bright and dark patches, ignores any possible localised changes in microphysical properties of the scattering particles or cloud-top altitudes.
We note however that a higher-order phase function correction scheme would not be supported by the relatively sparse and noisy dataset we have here.
An extensive analysis of how phase functions vary from dark to light regions on Venus, using the more sensitive 0-45° region of the phase function may be found elsewhere, see for example 
\citet{Petrova2015Icarus} and \citet{Shalygina2015PSS}.

\section {Analysis}
\label{analysis}

The goal of our analysis is to investigate whether any correlation can be found between the temperature structure derived from VeRa observations of a small area on Venus and the UV-brightness of that same area.
As discussed in the introduction, the UV contrasts are understood to originate in the upper clouds at around 70 $\pm$ 4 km altitude \citep{Crisp1986Icarus, SanchezLavega2008GRL, Ignatiev2009JGR, Lee2015PSS} where about half of the solar UV energy is deposited.
It would therefore be expected that any correlations preferentially occur in this altitude range.
The clouds are vertically extended with a scale height of several kilometres and therefore the deposition of energy is expected to happen over a range of altitudes and not be sharply bound.
We systematically explore this question by comparing the average (UV) Radiance Factor Ratio (RFR) per orbit and the temperature as derived from VeRa radio occultation,
as well as RFR and the static stability in the region between 50 and 80 km altitude. 

\begin{figure}[h!]
   \centering
   \includegraphics[width=\hsize]{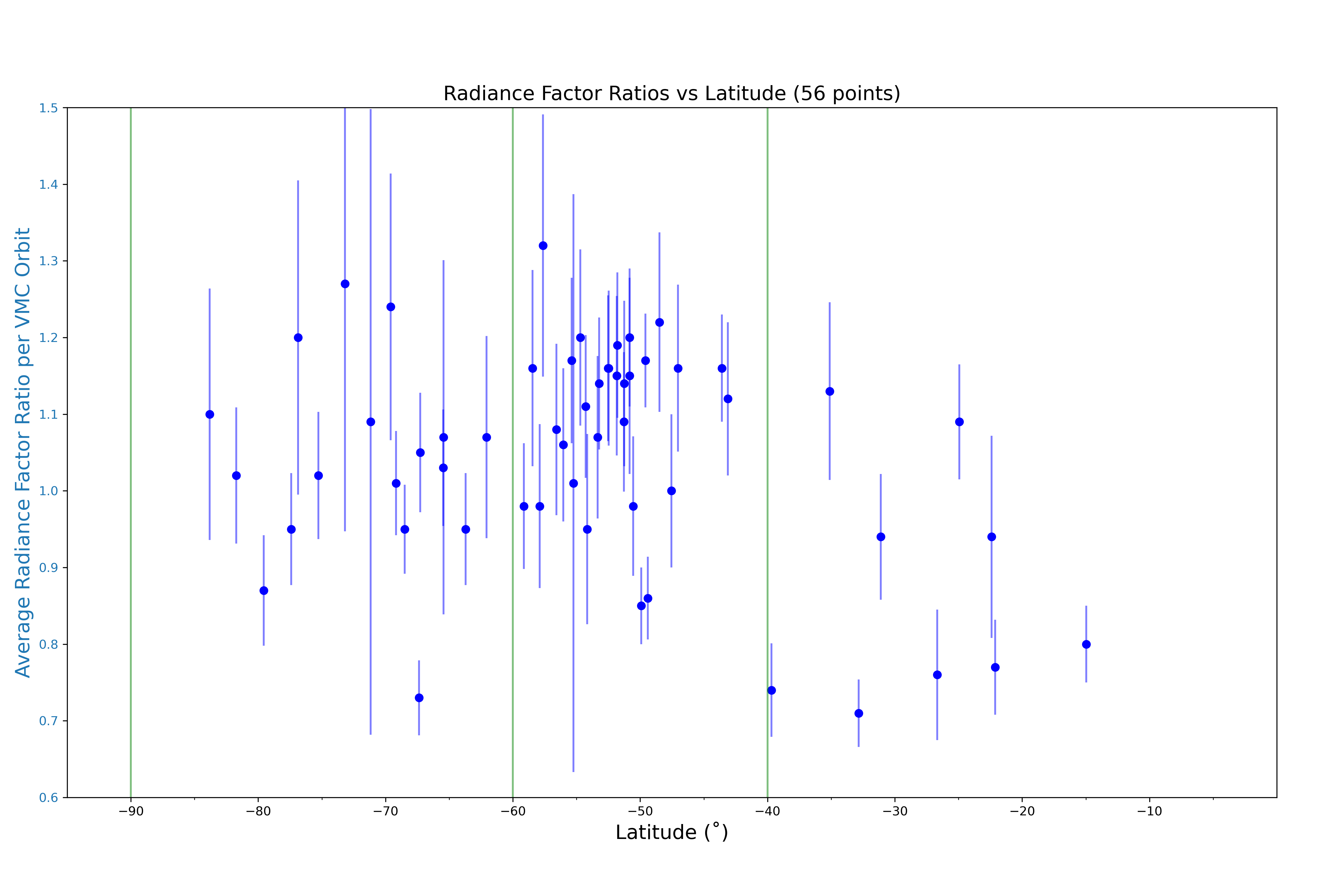}
      \caption{The average UV Radiance Factor Ratios of the VeRa sounding locations as a function of latitude of the VeRa sounding location. The green vertical lines indicate the division we make when doing the statistical analysis in terms of latitude, to try minimise the effect of changes with latitude of the UV-brightness.}
         \label{figure7}
   \end{figure}

It is well known that the temperature and the upper cloud structure on Venus show variations with latitude.
Not all of this variation has been explained however \citep{Lee2015PSS}.
Also, the overall UV-brightness changes with latitude.
In Fig.~\ref{figure7} we show the average Radiance Factor Ratios derived from our selection of images (Table \ref{table1}) versus latitude. 

From more exhaustive analysis it is known that the highest UV-brightness is typically found at mid-latitudes (-40$^{\circ}$ through -60$^{\circ}$), and the darkest regions are found at low latitudes 
\citep{Lee2015Icarus}.
We can see this trend  in Fig.~\ref{figure7}.

\begin{figure}[h!]
   \centering
   \includegraphics[width=\hsize]{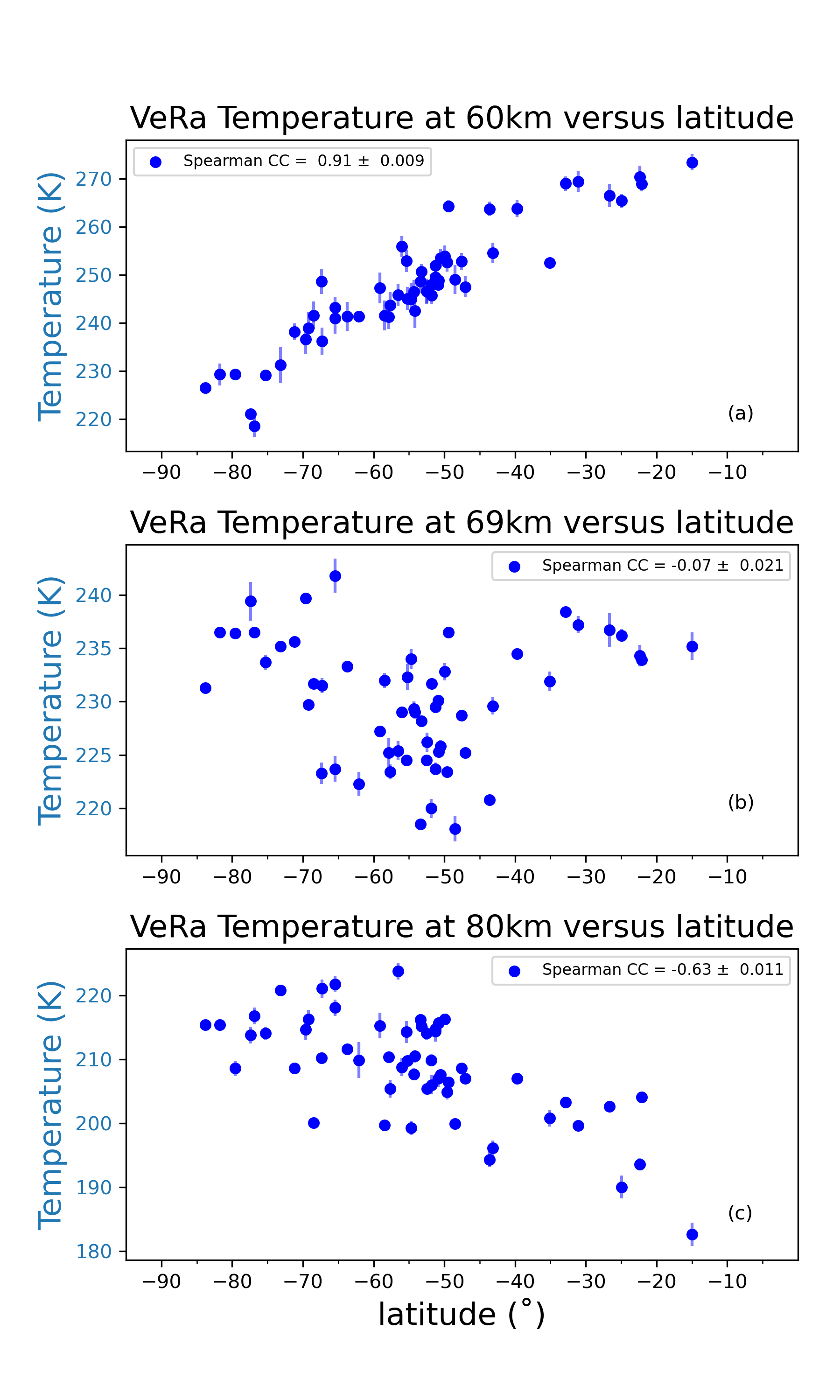}
      \caption{The temperature derived from 56 VeRa soundings (Table \ref{table1}) as a function of latitude at 60, 69 and 80 km altitude. There is a clear correlation at the lower and higher altitudes and no clear correlation in between. This is also clearly seen in Fig.~\ref{figure9} where the Spearman ranked and Pearson correlation coefficients are shown as a function of altitude. When there there is a correlation it seems to be fairly well described by a line, the reason for which is not obvious.}
         \label{figure8}
   \end{figure}

In Fig.~\ref{figure8} we present the VeRa temperatures at 60, 69 and 80 km altitude as a function of latitude from the 56 temperature profiles (Table \ref{table1}).

\begin{figure}[h!]
   \centering
   \includegraphics[width=\hsize]{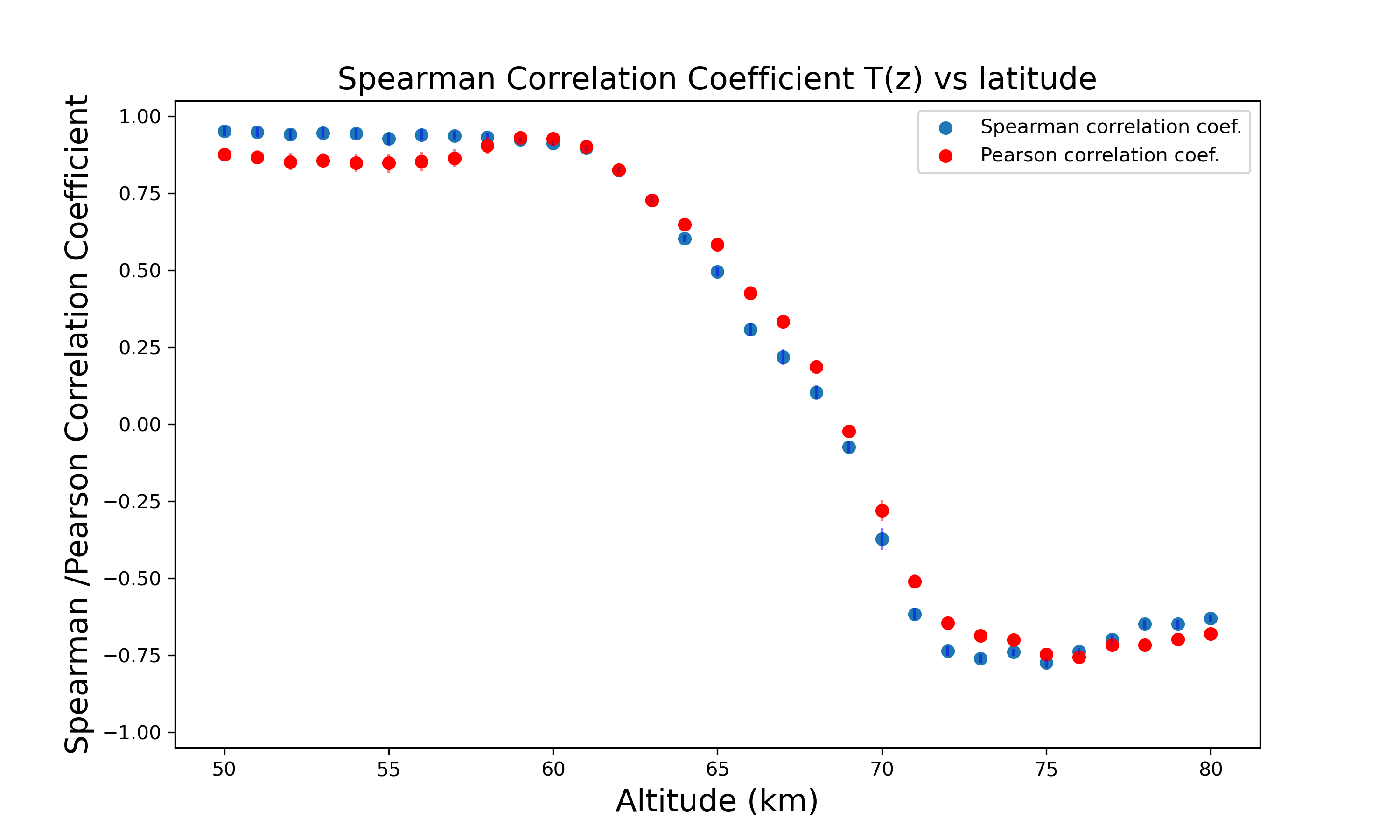}
      \caption{Pearson correlation coefficient (red) and Spearman ranked correlation coefficient (blue) for temperature as a function of latitude at altitudes between 50 and 80km, see also Fig.~\ref{figure8}. At the lower and higher altitude the correlation between temperature and latitude seems to be fairly well described with a line, the reason for which is not obvious.}
         \label{figure9}
   \end{figure}

This latitudinal variation echoes what has been previously reported for the full VeRa dataset by \citet{Tellmann2009JGR}, as well as from IR sounding \citep{Zasova2007PSS}:
 at higher altitudes the poles are warmer than the equator compared to lower altitudes, around the cloud tops there is a more equal situation.
 It is interesting to get an idea of the correlation between temperature and latitude without assuming any specific function.
 A suitable way to evaluate this is to calculate the Spearman ranked correlation coefficient. 
 This coefficient is a measure of how monotonically correlated the data is by evaluating the linearity of the ranked version of the data.
 The Spearman ranked correlation coefficient takes values between -1 and +1: the closer to (-)+1 the stronger is (anti)correlation, whereas around zero there is no correlation.
 In Fig.~\ref{figure9} we presnet the Spearman ranked correlation coefficient in blue as a function of altitude as derived from our data point (Figures 8a-c).

It can be seen that there is strong correlation for altitudes 50 - 60 km, strong anti-correlation for 70 - 80 km and weak or no correlation for 65 - 70 km.
From visual inspection of the examples in Fig.~\ref{figure8} a linear correlation seems to exist, though it is unclear why the correlation would be linear.
The Pearson correlation coefficient is a measure of the degree of linearity.
It is the covariance of the two variables divided by the product of their standard deviations.
Note that the Spearman ranked correlation coefficient is the Pearson correlation coefficient of the ranked data: while using the Pearson’s correlation we evaluate linear relationships, with the Spearman’s correlation we evaluate monotonic relationships, linear or not.
We calculate and show this coefficient in red in Fig.~\ref{figure9}.
We will use this to correct for the latitudinal variation as explained in the next section.

We decided to do our analysis in latitude bins in an effort to minimise latitudinal-variation effects.
From Fig.~\ref{figure7} and Fig.~\ref{figure9}, and from considering the cloud top altitude changes that affect the radiative balance \citep{Lee2015PSS} we decided on three latitude bins:
low latitudes between 0$^{\circ}$ and -40$^{\circ}$ (9 data points), mid latitudes between -40$^{\circ}$ and -60$^{\circ}$ (30 data points) and high latitudes between -60$^{\circ}$ and -90$^{\circ}$ (17 data points). 
These are indicated in Fig.~\ref{figure7} with the vertical green lines.

\subsection{UV-brightness versus Temperature correlation}
\label{uvbrightnessvstemperaturecorrelation}

\begin{figure}[h!]
   \centering
   \includegraphics[width=\hsize]{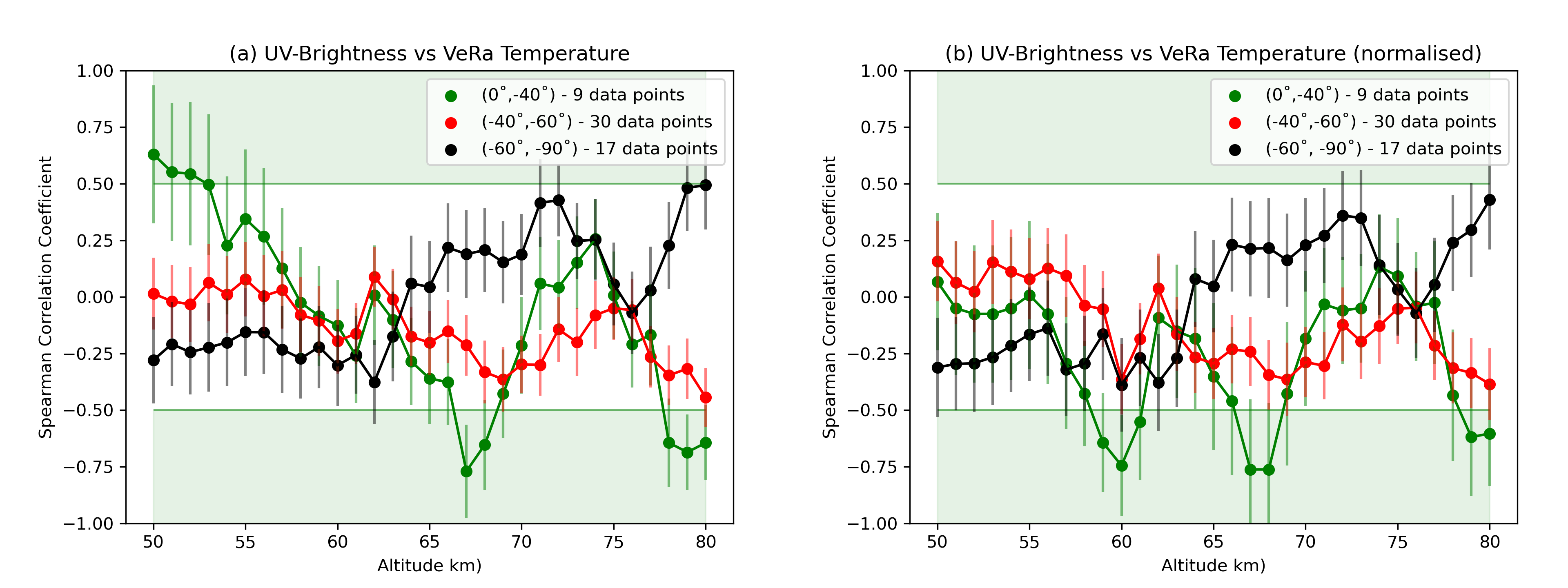}
      \caption{The Spearman ranked correlation coefficients for UV Radiance Factor Ratio as a function of (a) temperature and (b) normalised temperature, at levels between 50 and 80 km altitude for three latitude bins. Normalisation of the temperature is done to compensate for changes of the temperature with latitude that is not fully accounted for by dividing the data in three latitude bins. The largest effect is at the lower and higher altitudes, where the change of temperature with latitude is strongest (Fig.~\ref{figure8}, Fig.~\ref{figure9}, Fig.~\ref{figure11ab}.}
         \label{figure10ab}
   \end{figure}

\begin{figure}[h!]
   \centering
   \includegraphics[width=\hsize]{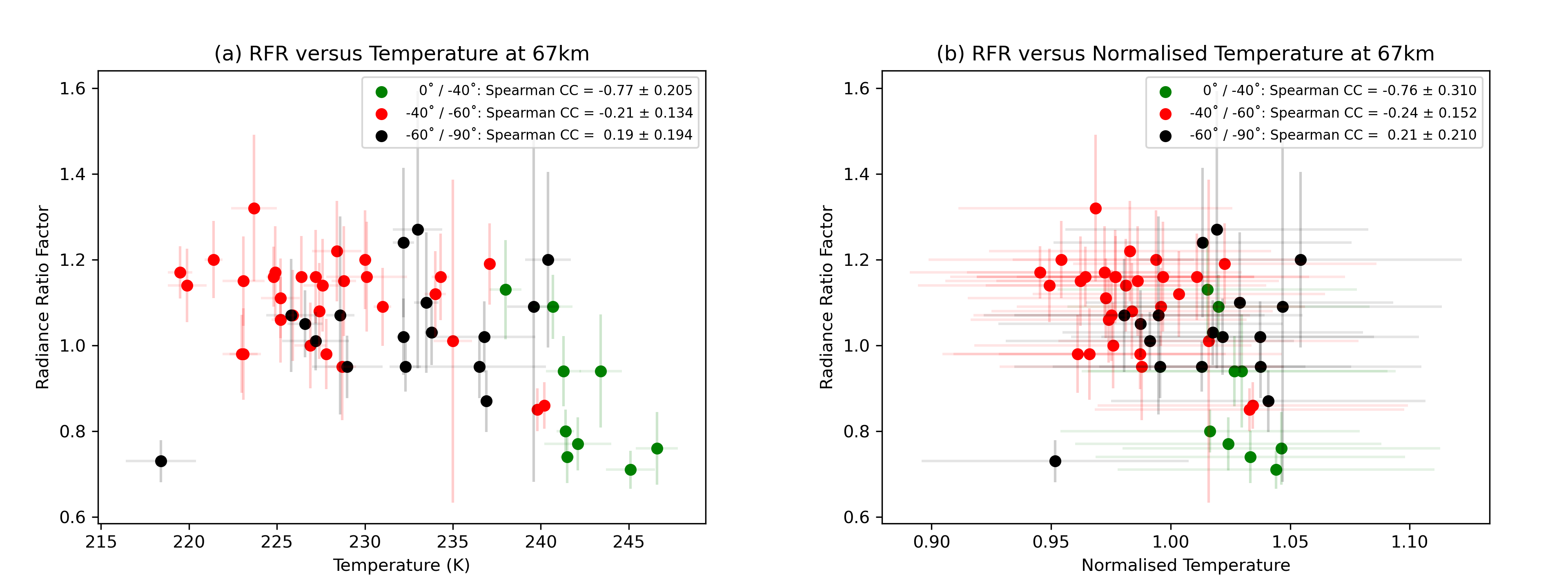}
      \caption{The Radiance Factor Ratio versus (a) temperature  and (b) normalised temperature at 67 km altitude. Normalisation is done relative to a linear least square fit to the temperature as a function of latitude at each level between 50 and 80 km altitude. See Fig.~\ref{figure8} and Fig.~\ref{figure9}.}
         \label{figure11ab}
   \end{figure}

We calculate the Spearman ranked correlation coefficients for the RFR versus the VeRa temperature at altitude levels between 50 and 80 km in each of the three latitude bins.
We estimate the uncertainties in the coefficients by running 1000 experiments changing the RFR and temperature values by adding random values from a Gaussian distribution that is characterised by the average and standard deviation
of each RFR and temperature values.
The result is shown in Fig.~\ref{figure10ab}a and an example of RFR versus VeRa temperature at 67 km altitude in Fig.~\ref{figure11ab}a.
The green areas in the figure indicate moderate to strong correlation.
A negative Spearman ranked correlation coefficient means that the UV-brightness decreases as the temperature increases.

As mentioned above and as can be seen from Fig.~\ref{figure8} and Fig.~\ref{figure9}, the temperatures show little variation with latitude in the 65-70 km altitude range, but a stronger variation at other altitudes. 
The latitude-binning process does not fully account for the effect of these temperature variations.
One way to address this would be to normalise the temperatures with respect to a linear model estimate of the temperature variation with latitude.
The evaluation of the Pearson correlation coefficient for the variation of the temperature with latitude seems to validate that a least square fitted line is a good option (Fig.~\ref{figure9}).
An example of RFR as a function of normalised temperature is shown for 67m altitude in Fig.~\ref{figure11ab}b.
Normalising the temperatures to such a model for each altitude, and performing the correlation analysis results in a correlation plot as shown in Fig.~\ref{figure10ab}b.
As expected, the structure and Spearman ranked correlation coefficient values remain very similar in the 65-70 km altitude range, but adjust to a more or a lesser degree at the other altitudes.

Despite our efforts to reduce the data as accurately and rigorously as possible, the data is rather noisy which leads to quite some scatter.
Yet we can identify some features we feel confident about.
The most obvious is that there seems to be a stronger tendency for anti-correlation between the UV-brightness and the temperature in the low and mid latitude bins in the 65 - 70 km altitude range: 
in these regions UV-dark areas are a few degrees hotter than UV-bright regions.
This could be a direct detection of the UV-energy deposition in this altitude range and hence it is instructive to examine how this temperature difference compares with theoretical expectations. 
\citet[Figure 11]{Crisp1986Icarus} and \citet{Tomasko1985AdSpR} present global mean heating rates for high and low-UV absorber cases, defined primarily from analysis of Pioneer Venus probe data, which have higher and lower heating rates respectively.
We can take these model-extremes as proxies for the variation we observe in the UV Radiance Factor Ratio.
In the 65-70 km altitude range the difference between the global mean heating rates from the two models is on the order of 2-4 K/(Earth)day.
The heating rate at the sub-solar point can be estimated to be a factor of 4 higher than the global mean heating rate (area of sphere versus area of circle).
If a UV-dark or UV-bright feature at the equator is sustained for a 2 Earth days, the time it takes to move from dawn to dusk at these altitudes, then the average heating rate over the course of 12h local solar time and changing solar incidence angle will be a factor of $\frac{2}{\pi}$ the sub-solar heating rate.
This results in a total factor of about 5 (4 times 2 (Earth days) times $\frac{2}{\pi}$).
Hence, the temperature difference between high-UV absorber and low-UV-absorber calculated from the Crisp model heating rates will be on the order of 10-20 K.
This is the calculated value at the equator, and would be correspondingly reduced by a factor of cosine(latitude) for other latitudes; if averaged over the latitude band of 0 - 40 degrees, the calculated temperature difference would be less by 10 percent or so, on the order of 9-18K.
In Fig.~\ref{figure11ab}a, where the UV Radiance Ratio Factor versus temperature at 67 km altitude is shown, the green points (low latitudes, showing anti-correlation, see Fig.~\ref{figure10ab}) span a range of about 10 K, which is most certainly in the same ballpark as the model estimated value.
This agreement between observed values and model predictions is encouraging, particularly given that this is observed only in the low latitude band (where solar heating would be expected to be most significant) and because the altitude range where the heating is observed also corresponds to that in Crisp’s model.
However, this cannot be considered a robust confirmation due to the scatter in the observational data.

  \begin{figure}[h!]
   \centering
   \includegraphics[width=\hsize]{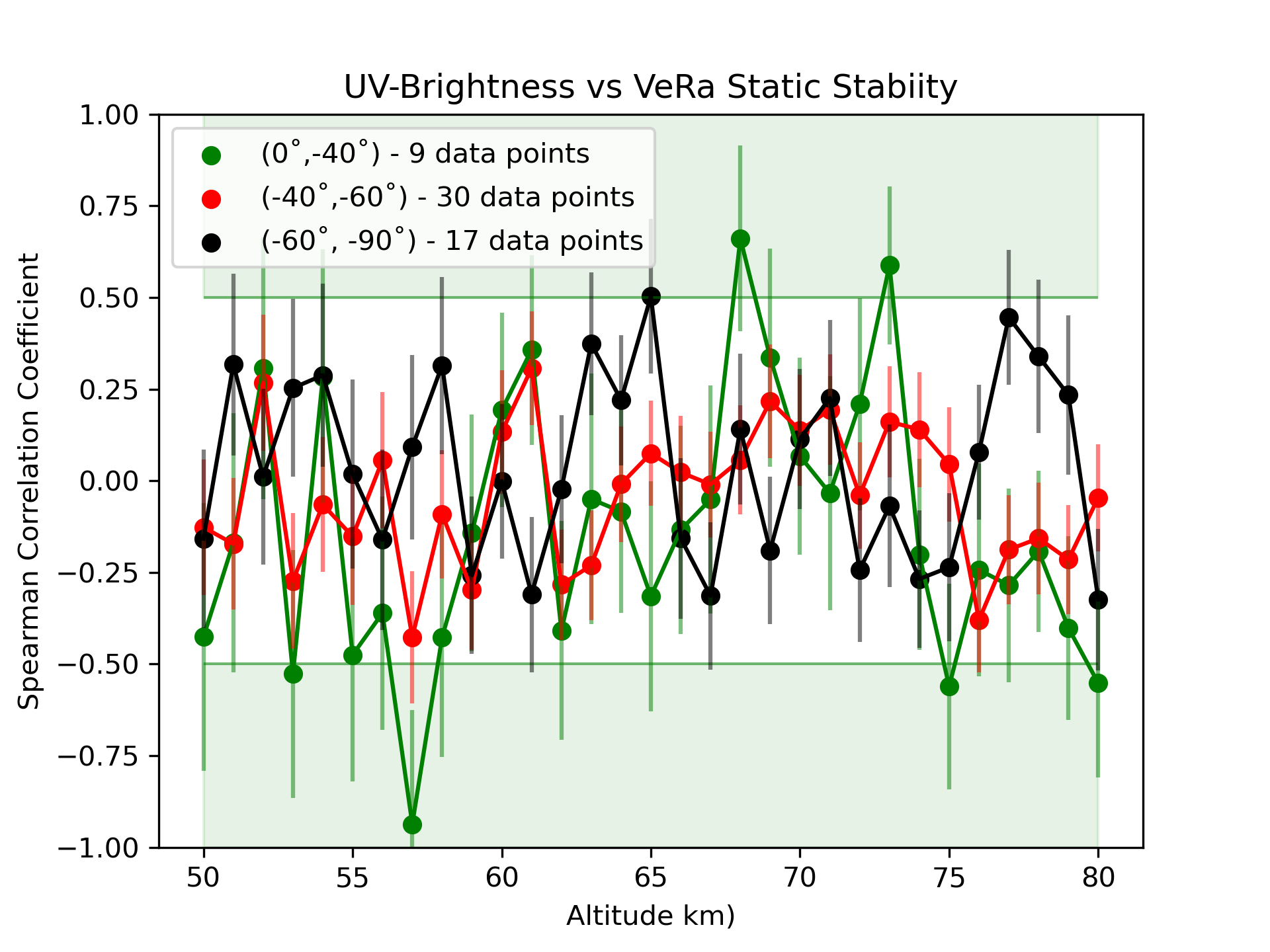}
      \caption{The Spearman ranked correlation coefficient for Radiance Factor Ratio as a function of static stability in the 50-80 km altitude range. The results present quite some uncertainty and scatter and there is no evidence for any trends.}
         \label{figure12}
   \end{figure}

\subsection{UV Brightness – stability correlations}
\label{uvbrightnessvsstabilitycorrelation}

We also analysed any possible correlation between the UV-brightness and the static stability.
We followed a similar process to the analysis on the correlation with temperature presented in the previous section.
The difference is that there is no variation of static stability with latitude in the 50 to 80 km altitude range and hence no need to perform any normalisation correction.
The Spearman ranked correlation coefficient for the RFR versus static stability for each altitude level is shown in Fig.~\ref{figure12}.
This result is even more scattered than for the temperature comparison and we find is no clear evidence for any significant correlation or trends.

\section{Conclusion}
\label{conclusion}

We have analysed the possibility of correlation between the UV-brightness and the temperature structure in the atmosphere using unique data from Venus Express.
On the one hand these data are measurements of the temperature structure from radio occultation data (VeRa experiment) in very small areas on Venus,
on the other hand they are UV-images of the same spot up to 11 hours before and a few hours after the radio occultation experiment (VMC instrument).
This type of analysis has not been presented before, as no such data exists from earlier missions.

We have reduced the data taking into account all sources of uncertainty.
Though the final result is still rather noisy, we find that there is a hint of detection of an anti-correlation between the atmospheric temperature in the 65-70 km altitude range and the UV-brightness in the corresponding region in the low to mid latitude range (latitude lower than -60$^{\circ}$).
No such correlation seems to be present for the polar latitudes.
From theoretical radiative forcing studies and in-situ descent probe measurements of the net radiative flux, it is known that the solar UV-energy is deposited in this altitude range.
The temperature difference between UV-dark and UV-bright regions, and its altitude dependence, is similar to that expected from increased solar heat absorption in UV-dark regions, approximately matching expectations.
Our result also provides possible vertical location of the solar energy deposition in the 65-70 km altitude range, implying that the UV-to-blue-absorber may exist in this vertical layer \citep{Tomasko1985AdSpR, Crisp1986Icarus, Lee2015PSS, Lee2021GRL} rather than being well-mixed through the upper haze layer \citep{Molaverdikhani2012Icarus}.
The assumption of a well-mixed unknown UV-to-blue-absorber above the cloud top level was also used for analysis of other spectral data  \citep{PerezHoyos2018JGR, Marcq2020Icarus}, while its impacts on solar heating have not yet been quantified.
Such vertical distribution may change over time \citet{Lee2015Icarus}, but since there is no similar data set for other earlier missions a possible temporal variation of heating altitudes cannot be done at this moment.

The present analysis has not shown any correlation between UV brightness and stability.
A correlation between UV brightness and stability could indicate that convective overturning is bringing UV-to-blue-absorber to the cloud-tops; conversely, an anti-correlation could indicate that convective overturning is bringing cloud-forming volatiles to the cloud-tops.
A further study may be conducted targeting static stability, using very fine vertical resolution data of radio occultation measurements similar to what \citet{Imamura2018JGR} achieved.

Although the present analysis has not reached robust conclusions using Venus Express data, the techniques shown here could be applied to larger datasets from future missions.

\begin{acknowledgements}

MRS thanks Emmanuel Marcq for sharing VIRTIS - SPICAV results and constructive discussion.

YJL was supported by the Institute for Basic Science (IBS-R035-C1).

A full note book and all the (Python) scripts created and used for this study can be found at the open Github repository: https://github.com/PleaseStateTheNatureOfYourInquiry/VenusResearchWorkBook 
(see also link to the ReadTheDocs-formatted note book at the end of the README section).

All images and temperature profiles used in this analysis are available at ...

The resulting wind tracking vectors for ultraviolet (365 nm) images obtained by the Venus Monitoring Camera (VMC) on board Venus Express for 18 orbits from the South Polar Dynamics Campaign (orbits numbers 2778-2811) are available in Mendeley Data \citep{Khatuntsev2024Mendeley} 

\end{acknowledgements}

\bibliographystyle{aa} 
\bibliography{venusbibtex.bib} 

\begin{thebibliography}{42}
\expandafter\ifx\csname natexlab\endcsname\relax\def\natexlab#1{#1}\fi

\bibitem[{{Akiba} {et~al.}(2021){Akiba}, {Taguchi}, {Fukuhara}, {Imamura},
  {Kouyama}, \& {Sato}}]{Akiba2021JGR}
{Akiba}, M., {Taguchi}, M., {Fukuhara}, T., {et~al.} 2021, Journal of
  Geophysical Research (Planets), 126, e06808

\bibitem[{{Barstow} {et~al.}(2012){Barstow}, {Tsang}, {Wilson}, {Irwin},
  {Taylor}, {McGouldrick}, {Drossart}, {Piccioni}, \&
  {Tellmann}}]{Barstow2012Icarus}
{Barstow}, J.~K., {Tsang}, C.~C.~C., {Wilson}, C.~F., {et~al.} 2012, \icarus,
  217, 542

\bibitem[{{Bevington} \& {Robinson}(2003)}]{Bevington2003}
{Bevington}, P.~R. \& {Robinson}, D.~K. 2003, {Data reduction and error
  analysis for the physical sciences}

\bibitem[{{Coblentz} \& {Lampland}(1925)}]{Coblentz1925}
{Coblentz}, W. \& {Lampland}, C. 1925, "Journal of the Franklin Inst", 119, 785

\bibitem[{{Cottini} {et~al.}(2015){Cottini}, {Ignatiev}, {Piccioni}, \&
  {Drossart}}]{Cottini2015PSS}
{Cottini}, V., {Ignatiev}, N.~I., {Piccioni}, G., \& {Drossart}, P. 2015,
  \planss, 113, 219

\bibitem[{{Crisp}(1986)}]{Crisp1986Icarus}
{Crisp}, D. 1986, \icarus, 67, 484

\bibitem[{{Ekonomov} {et~al.}(1984){Ekonomov}, {Moroz}, {Moshkin}, {Gnedykh},
  {Golovin}, \& {Crigoryev}}]{Ekonomov1984Nature}
{Ekonomov}, A.~P., {Moroz}, V.~I., {Moshkin}, B.~E., {et~al.} 1984, \nat, 307,
  345

\bibitem[{{Esposito} {et~al.}(1997){Esposito}, {Bertaux}, {Krasnopolsky},
  {Moroz}, \& {Zasova}}]{Esposito1997VenusII}
{Esposito}, L.~W., {Bertaux}, J.~L., {Krasnopolsky}, V., {Moroz}, V.~I., \&
  {Zasova}, L.~V. 1997, in Venus II: Geology, Geophysics, Atmosphere, and Solar
  Wind Environment, ed. S.~W. {Bougher}, D.~M. {Hunten}, \& R.~J. {Phillips},
  415

\bibitem[{{Frandsen} {et~al.}(2016){Frandsen}, {Wennberg}, \&
  {Kjaergaard}}]{Frandsen2016GRL}
{Frandsen}, B.~N., {Wennberg}, P.~O., \& {Kjaergaard}, H.~G. 2016, \grl, 43,
  11,146

\bibitem[{{Herrmann} {et~al.}(2014){Herrmann}, {H{\"a}usler}, {P{\"a}tzold},
  {Tellmann}, \& {Oschlisniok}}]{Herrmann2014EPSC}
{Herrmann}, M., {H{\"a}usler}, B., {P{\"a}tzold}, M., {Tellmann}, S., \&
  {Oschlisniok}, J. 2014, in European Planetary Science Congress, Vol.~9,
  EPSC2014--125

\bibitem[{{Ignatiev} {et~al.}(2009){Ignatiev}, {Titov}, {Piccioni}, {Drossart},
  {Markiewicz}, {Cottini}, {Roatsch}, {Almeida}, \& {Manoel}}]{Ignatiev2009JGR}
{Ignatiev}, N.~I., {Titov}, D.~V., {Piccioni}, G., {et~al.} 2009, Journal of
  Geophysical Research (Planets), 114, E00B43

\bibitem[{{Imamura} {et~al.}(2018){Imamura}, {Miyamoto}, {Ando}, {H{\"a}usler},
  {P{\"a}tzold}, {Tellmann}, {Tsuda}, {Aoyama}, {Murata}, {Takeuchi},
  {Yamazaki}, {Toda}, \& {Tomiki}}]{Imamura2018JGR}
{Imamura}, T., {Miyamoto}, M., {Ando}, H., {et~al.} 2018, Journal of
  Geophysical Research (Planets), 123, 2151

\bibitem[{{Khatuntsev} \& {Patsaeva}(2024)}]{Khatuntsev2024Mendeley}
{Khatuntsev}, I.~V. \& {Patsaeva}, M. 2024, {VMC wind measurements during the
  South Pole Dynamics Campaign}

\bibitem[{{Khatuntsev} {et~al.}(2013){Khatuntsev}, {Patsaeva}, {Titov},
  {Ignatiev}, {Turin}, {Limaye}, {Markiewicz}, {Almeida}, {Roatsch}, \&
  {Moissl}}]{Khatuntsev2013Icarus}
{Khatuntsev}, I.~V., {Patsaeva}, M.~V., {Titov}, D.~V., {et~al.} 2013, \icarus,
  226, 140

\bibitem[{{Kouyama} {et~al.}(2017){Kouyama}, {Imamura}, {Taguchi}, {Fukuhara},
  {Sato}, {Yamazaki}, {Futaguchi}, {Murakami}, {Hashimoto}, {Ueno}, {Iwagami},
  {Takagi}, {Takagi}, {Ogohara}, {Kashimura}, {Horinouchi}, {Sato}, {Yamada},
  {Yamamoto}, {Ohtsuki}, {Sugiyama}, {Ando}, {Takamura}, {Yamada}, {Satoh}, \&
  {Nakamura}}]{Kouyama2017GRL}
{Kouyama}, T., {Imamura}, T., {Taguchi}, M., {et~al.} 2017, \grl, 44, 12,098

\bibitem[{{Lee} {et~al.}(2020){Lee}, {Garc{\'\i}a Mu{\~n}oz}, {Imamura},
  {Yamada}, {Satoh}, {Yamazaki}, \& {Watanabe}}]{Lee2020Nature}
{Lee}, Y.~J., {Garc{\'\i}a Mu{\~n}oz}, A., {Imamura}, T., {et~al.} 2020, Nature
  Communications, 11, 5720

\bibitem[{{Lee} {et~al.}(2021){Lee}, {Garc{\'\i}a Mu{\~n}oz}, {Yamazaki},
  {Yamada}, {Watanabe}, \& {Encrenaz}}]{Lee2021GRL}
{Lee}, Y.~J., {Garc{\'\i}a Mu{\~n}oz}, A., {Yamazaki}, A., {et~al.} 2021, \grl,
  48, e90577

\bibitem[{{Lee} {et~al.}(2015{\natexlab{a}}){Lee}, {Imamura}, {Schr{\"o}der},
  \& {Marcq}}]{Lee2015Icarus}
{Lee}, Y.~J., {Imamura}, T., {Schr{\"o}der}, S.~E., \& {Marcq}, E.
  2015{\natexlab{a}}, \icarus, 253, 1

\bibitem[{{Lee} {et~al.}(2019){Lee}, {Jessup}, {Perez-Hoyos}, {Titov},
  {Lebonnois}, {Peralta}, {Horinouchi}, {Imamura}, {Limaye}, {Marcq}, {Takagi},
  {Yamazaki}, {Yamada}, {Watanabe}, {Murakami}, {Ogohara}, {McClintock},
  {Holsclaw}, \& {Roman}}]{Lee2019AJ}
{Lee}, Y.~J., {Jessup}, K.-L., {Perez-Hoyos}, S., {et~al.} 2019, \aj, 158, 126

\bibitem[{{Lee} {et~al.}(2015{\natexlab{b}}){Lee}, {Titov}, {Ignatiev},
  {Tellmann}, {P{\"a}tzold}, \& {Piccioni}}]{Lee2015PSS}
{Lee}, Y.~J., {Titov}, D.~V., {Ignatiev}, N.~I., {et~al.} 2015{\natexlab{b}},
  \planss, 113, 298

\bibitem[{{Marcq} {et~al.}(2019){Marcq}, {Baggio}, {Lef{\`e}vre},
  {Stolzenbach}, {Montmessin}, {Belyaev}, {Korablev}, \&
  {Bertaux}}]{Marcq2019Icarus}
{Marcq}, E., {Baggio}, L., {Lef{\`e}vre}, F., {et~al.} 2019, \icarus, 319, 491

\bibitem[{{Marcq} {et~al.}(2020){Marcq}, {Lea Jessup}, {Baggio}, {Encrenaz},
  {Lee}, {Montmessin}, {Belyaev}, {Korablev}, \& {Bertaux}}]{Marcq2020Icarus}
{Marcq}, E., {Lea Jessup}, K., {Baggio}, L., {et~al.} 2020, \icarus, 335,
  113368

\bibitem[{{Markiewicz} {et~al.}(2007){Markiewicz}, {Titov}, {Ignatiev},
  {Keller}, {Crisp}, {Limaye}, {Jaumann}, {Moissl}, {Thomas}, {Esposito},
  {Watanabe}, {Fiethe}, {Behnke}, {Szemerey}, {Michalik}, {Perplies},
  {Wedemeier}, {Sebastian}, {Boogaerts}, {Hviid}, {Dierker}, {Osterloh},
  {B{\"o}ker}, {Koch}, {Michaelis}, {Belyaev}, {Dannenberg}, {Tschimmel},
  {Russo}, {Roatsch}, \& {Matz}}]{Markiewicz2007PSS}
{Markiewicz}, W.~J., {Titov}, D.~V., {Ignatiev}, N., {et~al.} 2007, \planss,
  55, 1701

\bibitem[{{Molaverdikhani} {et~al.}(2012){Molaverdikhani}, {McGouldrick}, \&
  {Esposito}}]{Molaverdikhani2012Icarus}
{Molaverdikhani}, K., {McGouldrick}, K., \& {Esposito}, L.~W. 2012, \icarus,
  217, 648

\bibitem[{{Patsaeva} {et~al.}(2015){Patsaeva}, {Khatuntsev}, {Patsaev},
  {Titov}, {Ignatiev}, {Markiewicz}, \& {Rodin}}]{Patsaeva2015PSS}
{Patsaeva}, M.~V., {Khatuntsev}, I.~V., {Patsaev}, D.~V., {et~al.} 2015,
  \planss, 113, 100

\bibitem[{{P{\"a}tzold} {et~al.}(2007){P{\"a}tzold}, {H{\"a}usler}, {Bird},
  {Tellmann}, {Mattei}, {Asmar}, {Dehant}, {Eidel}, {Imamura}, {Simpson}, \&
  {Tyler}}]{Patzold2007Nature}
{P{\"a}tzold}, M., {H{\"a}usler}, B., {Bird}, M.~K., {et~al.} 2007, \nat, 450,
  657

\bibitem[{{P{\'e}rez-Hoyos} {et~al.}(2018){P{\'e}rez-Hoyos},
  {S{\'a}nchez-Lavega}, {Garc{\'\i}a-Mu{\~n}oz}, {Irwin}, {Peralta},
  {Holsclaw}, {McClintock}, \& {Sanz-Requena}}]{PerezHoyos2018JGR}
{P{\'e}rez-Hoyos}, S., {S{\'a}nchez-Lavega}, A., {Garc{\'\i}a-Mu{\~n}oz}, A.,
  {et~al.} 2018, Journal of Geophysical Research (Planets), 123, 145

\bibitem[{{Petrova} {et~al.}(2015){Petrova}, {Shalygina}, \&
  {Markiewicz}}]{Petrova2015Icarus}
{Petrova}, E.~V., {Shalygina}, O.~S., \& {Markiewicz}, W.~J. 2015, \icarus,
  260, 190

\bibitem[{{Pollack} {et~al.}(1980){Pollack}, {Toon}, {Whitten}, {Boese},
  {Ragent}, {Tomasko}, {Eposito}, {Travis}, \& {Wiedman}}]{Pollack1980JGR}
{Pollack}, J.~B., {Toon}, O.~B., {Whitten}, R.~C., {et~al.} 1980, \jgr, 85,
  8141

\bibitem[{{Ross}(1928)}]{Ross1928ApJ}
{Ross}, F.~E. 1928, \apj, 68, 57

\bibitem[{{S{\'a}nchez-Lavega} {et~al.}(2008){S{\'a}nchez-Lavega}, {Hueso},
  {Piccioni}, {Drossart}, {Peralta}, {P{\'e}rez-Hoyos}, {Wilson}, {Taylor},
  {Baines}, {Luz}, {Erard}, \& {Lebonnois}}]{SanchezLavega2008GRL}
{S{\'a}nchez-Lavega}, A., {Hueso}, R., {Piccioni}, G., {et~al.} 2008, \grl, 35,
  L13204

\bibitem[{{Seiff} {et~al.}(1980){Seiff}, {Kirk}, {Young}, {Blanchard},
  {Findlay}, {Kelly}, \& {Sommer}}]{Seiff1980JGR}
{Seiff}, A., {Kirk}, D.~B., {Young}, R.~E., {et~al.} 1980, \jgr, 85, 7903

\bibitem[{{Shalygina} {et~al.}(2015){Shalygina}, {Petrova}, {Markiewicz},
  {Ignatiev}, \& {Shalygin}}]{Shalygina2015PSS}
{Shalygina}, O.~S., {Petrova}, E.~V., {Markiewicz}, W.~J., {Ignatiev}, N.~I.,
  \& {Shalygin}, E.~V. 2015, \planss, 113, 135

\bibitem[{{Svedhem} {et~al.}(2007){Svedhem}, {Titov}, {McCoy}, {Lebreton},
  {Barabash}, {Bertaux}, {Drossart}, {Formisano}, {H{\"a}usler}, {Korablev},
  {Markiewicz}, {Nevejans}, {P{\"a}tzold}, {Piccioni}, {Zhang}, {Taylor},
  {Lellouch}, {Koschny}, {Witasse}, {Eggel}, {Warhaut}, {Accomazzo},
  {Rodriguez-Canabal}, {Fabrega}, {Schirmann}, {Clochet}, \&
  {Coradini}}]{Svedhem2007PSS}
{Svedhem}, H., {Titov}, D.~V., {McCoy}, D., {et~al.} 2007, \planss, 55, 1636

\bibitem[{{Tellmann} {et~al.}(2012){Tellmann}, {H{\"a}usler}, {Hinson},
  {Tyler}, {Andert}, {Bird}, {Imamura}, {P{\"a}tzold}, \&
  {Remus}}]{Tellmann2012Icarus}
{Tellmann}, S., {H{\"a}usler}, B., {Hinson}, D.~P., {et~al.} 2012, \icarus,
  221, 471

\bibitem[{{Tellmann} {et~al.}(2009){Tellmann}, {P{\"a}tzold}, {H{\"a}usler},
  {Bird}, \& {Tyler}}]{Tellmann2009JGR}
{Tellmann}, S., {P{\"a}tzold}, M., {H{\"a}usler}, B., {Bird}, M.~K., \&
  {Tyler}, G.~L. 2009, Journal of Geophysical Research (Planets), 114, E00B36

\bibitem[{{Titov} {et~al.}(2008){Titov}, {Taylor}, {Svedhem}, {Ignatiev},
  {Markiewicz}, {Piccioni}, \& {Drossart}}]{Titov2008Nature}
{Titov}, D.~V., {Taylor}, F.~W., {Svedhem}, H., {et~al.} 2008, \nat, 456, 620

\bibitem[{{Tomasko} {et~al.}(1985){Tomasko}, {Doose}, \&
  {Smith}}]{Tomasko1985AdSpR}
{Tomasko}, M.~G., {Doose}, L.~R., \& {Smith}, P.~H. 1985, Advances in Space
  Research, 5, 71

\bibitem[{{Tomasko} {et~al.}(1980){Tomasko}, {Doose}, {Smith}, \&
  {Odell}}]{Tomasko1980JGR}
{Tomasko}, M.~G., {Doose}, L.~R., {Smith}, P.~H., \& {Odell}, A.~P. 1980, \jgr,
  85, 8167

\bibitem[{{Toon} {et~al.}(1982){Toon}, {Turco}, \& {Pollack}}]{Toon1982Icarus}
{Toon}, O.~B., {Turco}, R.~P., \& {Pollack}, J.~B. 1982, \icarus, 51, 358

\bibitem[{{Wright}(1927)}]{Wright1927PASP}
{Wright}, W.~H. 1927, \pasp, 39, 220

\bibitem[{{Zasova} {et~al.}(2007){Zasova}, {Ignatiev}, {Khatuntsev}, \&
  {Linkin}}]{Zasova2007PSS}
{Zasova}, L.~V., {Ignatiev}, N., {Khatuntsev}, I., \& {Linkin}, V. 2007,
  \planss, 55, 1712

\end{thebibliography}

\begin {appendix}

\begin{table}
\caption{Venus Express orbits with VeRa radio occultation soundings}
\label{table1}
\centering
\begin{tabular}{c c c r c c c c}
\hline
OrbitID & Date & VeRa               & \# images & RFR.        & $\Delta$RFR & RFR         & $\Delta$RFR  \\
            &         & ($\theta, \phi$) &                  & (average) & (of average)  & (median) & (of median) \\

\hline
\hline
1191 & 2009-07-25 & -15$^{\circ}$, 143$^{\circ}$ & 9 & 0.80 & 0.049 & 0.82 & 0.031 \\
1193 & 2009-07-27 & -23$^{\circ}$, 148$^{\circ}$ & 24 & 0.94 & 0.132 & 0.94 & 0.031 \\
1201 & 2009-08-04 & -50$^{\circ}$, 170$^{\circ}$ & 26 & 1.17 & 0.060 & 1.15 & 0.040 \\
1204 & 2009-08-07 & -59$^{\circ}$, 178$^{\circ}$ & 30 & 0.98 & 0.082 & 0.96 & 0.027 \\
1207 & 2009-08-10 & -69$^{\circ}$, 187$^{\circ}$ & 25 & 0.95 & 0.057 & 0.93 & 0.015 \\
1513 & 2010-06-12 & -67$^{\circ}$, 121$^{\circ}$ & 34 & 0.73 & 0.049 & 0.72 & 0.010 \\
1521 & 2010-06-20 & -44$^{\circ}$, 139$^{\circ}$ & 40 & 1.16 & 0.069 & 1.16 & 0.018 \\
\hline
1748 & 2011-02-02 & -33$^{\circ}$, 072$^{\circ}$ & 36 & 0.71 & 0.043 & 0.71 & 0.019 \\
1753 & 2011-02-07 & -50$^{\circ}$, 085$^{\circ}$ & 35 & 0.86 & 0.053 & 0.87 & 0.012 \\
1758 & 2011-02-12 & -66$^{\circ}$, 096$^{\circ}$ & 33 & 1.03 & 0.075 & 1.03 & 0.016 \\
1901 & 2011-07-05 & -50$^{\circ}$, 123$^{\circ}$ & 18 & 0.85 & 0.048 & 0.83 & 0.013 \\
2063 & 2011-12-14 & -77$^{\circ}$, 017$^{\circ}$ & 24 & 1.20 & 0.205 & 1.21 & 0.058 \\
2065 & 2011-12-16 & -70$^{\circ}$, 026$^{\circ}$ & 22 & 1.24 & 0.173 & 1.27 & 0.041 \\
2068 & 2011-12-19 & -59$^{\circ}$, 036$^{\circ}$ & 14 & 1.16 & 0.127 & 1.15 & 0.038 \\
2069 & 2011-12-20 & -55$^{\circ}$, 039$^{\circ}$ & 15 & 1.20 & 0.114 & 1.19 & 0.035 \\
2071 & 2011-12-22 & -47$^{\circ}$, 045$^{\circ}$ & 14 & 1.16 & 0.108 & 1.14 & 0.034 \\
2072 & 2011-12-23 & -43$^{\circ}$, 048$^{\circ}$ & 16 & 1.12 & 0.100 & 1.11 & 0.029 \\
2074 & 2011-12-25 & -35$^{\circ}$, 054$^{\circ}$ & 8 & 1.13 & 0.115 & 1.15 & 0.044 \\
2300 & 2012-08-07 & -49$^{\circ}$, 357$^{\circ}$ & 6 & 1.22 & 0.116 & 1.21 & 0.059 \\
\hline
2452 & 2013-01-06 & -22$^{\circ}$, 035$^{\circ}$ & 7 & 0.77 & 0.062 & 0.77 & 0.039 \\
2453 & 2013-01-07 & -27$^{\circ}$, 038$^{\circ}$ & 10 & 0.76 & 0.085 & 0.75 & 0.048 \\
2454 & 2013-01-08 & -31$^{\circ}$, 041$^{\circ}$ & 7 & 0.94 & 0.082 & 0.93 & 0.037 \\
2456 & 2013-01-10 & -40$^{\circ}$, 047$^{\circ}$ & 5 & 0.74 & 0.061 & 0.73 & 0.033 \\
2460 & 2013-01-14 & -56$^{\circ}$, 058$^{\circ}$ & 7 & 1.06 & 0.099 & 1.04 & 0.045 \\
2630 & 2013-07-03 & -48$^{\circ}$, 341$^{\circ}$ & 14 & 1.00 & 0.100 & 0.96 & 0.020 \\
2638 & 2013-07-11 & -25$^{\circ}$, 002$^{\circ}$ & 11 & 1.09 & 0.075 & 1.10 & 0.054 \\
\hline
2776 & 2013-11-26 & -84$^{\circ}$, 323$^{\circ}$ & 9 & 1.10 & 0.164 & 1.08 & 0.067 \\
2777 & 2013-11-27 & -82$^{\circ}$, 336$^{\circ}$ & 8 & 1.02 & 0.088 & 1.02 & 0.035 \\
2778 & 2013-11-28 & -80$^{\circ}$, 344$^{\circ}$ & 9 & 0.87 & 0.071 & 0.88 & 0.031 \\
2779 & 2013-11-29 & -77$^{\circ}$, 349$^{\circ}$ & 8 & 0.95 & 0.072 & 0.94 & 0.028 \\
2780 & 2013-11-30 & -75$^{\circ}$, 354$^{\circ}$ & 9 & 1.02 & 0.082 & 1.01 & 0.033 \\
2781 & 2013-12-01 & -73$^{\circ}$, 358$^{\circ}$ & 13 & 1.27 & 0.323 & 1.07 & 0.246 \\
2782 & 2013-12-02 & -71$^{\circ}$, 002$^{\circ}$ & 17 & 1.09 & 0.408 & 0.99 & 0.031 \\
2783 & 2013-12-03 & -69$^{\circ}$, 005$^{\circ}$ & 10 & 1.01 & 0.067 & 1.02 & 0.037 \\
2784 & 2013-12-04 & -67$^{\circ}$, 008$^{\circ}$ & 12 & 1.05 & 0.077 & 1.04 & 0.026 \\
2785 & 2013-12-05 & -66$^{\circ}$, 011$^{\circ}$ & 16 & 1.07 & 0.231 & 0.97 & 0.057 \\
2786 & 2013-12-06 & -64$^{\circ}$, 014$^{\circ}$ & 11 & 0.95 & 0.072 & 0.93 & 0.027 \\
2787 & 2013-12-07 & -62$^{\circ}$, 016$^{\circ}$ & 9 & 1.07 & 0.131 & 1.04 & 0.071 \\
2790 & 2013-12-10 & -58$^{\circ}$, 024$^{\circ}$ & 12 & 1.32 & 0.171 & 1.26 & 0.048 \\
2792 & 2013-12-12 & -55$^{\circ}$, 028$^{\circ}$ & 9 & 1.01 & 0.377 & 1.05 & 0.039 \\
2793 & 2013-12-13 & -54$^{\circ}$, 030$^{\circ}$ & 21 & 0.95 & 0.124 & 0.91 & 0.021 \\
2794 & 2013-12-14 & -53$^{\circ}$, 032$^{\circ}$ & 18 & 1.14 & 0.086 & 1.12 & 0.036 \\
2795 & 2013-12-15 & -53$^{\circ}$, 034$^{\circ}$ & 23 & 1.16 & 0.101 & 1.14 & 0.024 \\
2796 & 2013-12-16 & -52$^{\circ}$, 036$^{\circ}$ & 20 & 1.19 & 0.096 & 1.16 & 0.026 \\
2797 & 2013-12-17 & -51$^{\circ}$, 038$^{\circ}$ & 19 & 1.14 & 0.109 & 1.12 & 0.035 \\
2798 & 2013-12-18 & -51$^{\circ}$, 040$^{\circ}$ & 16 & 1.20 & 0.090 & 1.17 & 0.032 \\
2802 & 2013-12-22 & -51$^{\circ}$, 047$^{\circ}$ & 19 & 0.98 & 0.091 & 0.96 & 0.046 \\
2803 & 2013-12-23 & -51$^{\circ}$, 049$^{\circ}$ & 21 & 1.15 & 0.128 & 1.16 & 0.034 \\
2804 & 2013-12-24 & -51$^{\circ}$, 050$^{\circ}$ & 20 & 1.09 & 0.091 & 1.07 & 0.038 \\
2805 & 2013-12-25 & -52$^{\circ}$, 052$^{\circ}$ & 18 & 1.15 & 0.104 & 1.14 & 0.027 \\
2806 & 2013-12-26 & -53$^{\circ}$, 054$^{\circ}$ & 21 & 1.16 & 0.095 & 1.16 & 0.047 \\
2807 & 2013-12-27 & -54$^{\circ}$, 055$^{\circ}$ & 21 & 1.07 & 0.106 & 1.04 & 0.028 \\
2808 & 2013-12-28 & -54$^{\circ}$, 057$^{\circ}$ & 21 & 1.11 & 0.092 & 1.10 & 0.027 \\
2809 & 2013-12-29 & -56$^{\circ}$, 059$^{\circ}$ & 21 & 1.17 & 0.107 & 1.16 & 0.027 \\
2810 & 2013-12-30 & -57$^{\circ}$, 060$^{\circ}$ & 19 & 1.08 & 0.112 & 1.03 & 0.025 \\
2811 & 2013-12-31 & -58$^{\circ}$, 062$^{\circ}$ & 21 & 0.99 & 0.119 & 0.97 & 0.031 \\
\hline
\hline
\end{tabular}
\end{table}

\end {appendix}

\end{document}